\documentclass[12pt,letterpaper]{article}
\usepackage{jheplikemod,tocloft,ifpdf,tocloft,rotating,amsmath,mathrsfs,multirow,mathtools,xcolor,cancel,tikz}
\usetikzlibrary{calc,decorations.markings,shapes.geometric}
\usepackage[backgroundcolor=green!5, linecolor=blue!60]{todonotes}

\definecolor{hteal}{rgb}{0.0,0.545,0.7451}
\definecolor{cut2}{rgb}{0.2353,0.2353,0.60}
\definecolor{cut1}{rgb}{0.60,0.0353,0.2353}
\definecolor{rindou1}{rgb}{0.4431,0.2862,0.7960}
\definecolor{rindou2}{rgb}{0.0078,0.1215,0.4392}
\definecolor{lapis}{rgb}{0.0.0470,0.2941,0.5568}
\definecolor{emerald}{rgb}{0.31, 0.78, 0.47}
\definecolor{pinegreen}{rgb}{0.0, 0.47, 0.44}
\definecolor{jade}{rgb}{0.0, 0.66, 0.42}
\definecolor{teal}{rgb}{0.0, 0.5, 0.5}
\definecolor{hblue}{rgb}{0,0,0.575}
\definecolor{hred}{rgb}{0.575,0.0,0.225}
\definecolor{hgreen}{rgb}{0.0,0.4,0.2}
\definecolor{hteal}{rgb}{0.0,0.545,0.7451}

\newcommand{\merge}{\raisebox{-1.5pt}{\scalebox{1.5}{$\otimes$}}}

\definecolor{nmhv}{rgb}{0.95,0.55,0.55}
\newcommand{\eps}{\delta}
\newcommand{\logsquare}[1]{\log^{\hspace{-0.25pt}2}\hspace{-1pt}(#1)}
\newcommand{\logsquarediv}{\logsquare{\eps}}
\newcommand{\logdiv}{\log(\eps)}

\definecolor{emphA}{rgb}{0.75,0.0,0.325}
\definecolor{emphB}{rgb}{0.2,0.605,0}
\definecolor{emphC}{rgb}{0,0,0.975}
\definecolor{ndotColor}{rgb}{0.65,0.25,0.25}

\definecolor{optLegColour}{rgb}{0.5,0.5,0.5}
\definecolor{legColour}{rgb}{0.35,0.35,0.35}
\definecolor{legLabelColour}{rgb}{0,0,0}
\definecolor{mhvblue}{rgb}{0.6,0.6,0.7765}
\definecolor{ampgrey}{rgb}{0.9,0.9,0.9}

\tikzset{int/.style={black,line width=\lineThickness,line cap=round,rounded corners=0.5pt}}
\tikzset{intH/.style={emphA,line width=\lineThickness,line cap=round,rounded corners=0.5pt}}
\tikzset{fullamp/.style={coordinate,minimum size=0.9*\ampSize,ball color=black!20,circle,text=white,inner sep=0}}
\tikzset{fullmhv/.style={coordinate,minimum size=1.2*\ampSize,ball color=mhvblue,circle,text=white,inner sep=0}}
\tikzset{fullmhvBar/.style={coordinate,minimum size=0.9*\ampSize,ball color=white,circle,text=white,inner sep=0}}
\tikzset{mhv/.style={fill=mhvblue,circle,draw=black,line width=\lineThickness,minimum size=0.5*\ampSize,text=white,inner sep=0}}
\tikzset{mhvBar/.style={fill=white,circle,draw=black,line width=\lineThickness,minimum size=0.5*\ampSize,text=white,inner sep=0}}
\tikzset{genAmp/.style={fill=ampgrey,circle,draw=black,line width=\lineThickness,minimum size=0.5*\ampSize,text=white,inner sep=0}}
\tikzset{genAmpFull/.style={coordinate,minimum size=0.9*\ampSize,ball color=ampgrey,circle,text=white,inner sep=0}}

\newcommand{\singleLeg}[2]{
\draw[int] #1--($#1+(#2:\extLegLen*1)$);
\node at #1 [ddot]{};
}
\newcommand{\singleLegLabelled}[3]{\node at #1 [ddot]{};\draw[int] #1--($#1+(#2:\extLegLen*1.25)$);
\node at ($#1+(#2:\labelDist*1.35)$)[]{{\footnotesize #3}};}

\newcommand{\generalCornerMassiveLabelled}[4]{\coordinate (aa0) at #1;
\coordinate (aa5) at ($#1+(#2+\legSpread*3:\extLegLen)$);
\coordinate (bb5) at ($#1+(#2-\legSpread*3:\extLegLen)$);
\coordinate (aa1) at ($(aa0)!0.4!(aa5)$);\coordinate (aa2) at ($(aa0)!0.55!(aa5)$);\coordinate (aa3) at ($(aa0)!0.7!(aa5)$);\coordinate (aa4) at ($(aa0)!0.85!(aa5)$);
\coordinate (bb1) at ($(aa0)!0.4!(bb5)$);\coordinate (bb2) at ($(aa0)!0.55!(bb5)$);\coordinate (bb3) at ($(aa0)!0.7!(bb5)$);\coordinate (bb4) at ($(aa0)!0.85!(bb5)$);
\fill[massiveCornerColour] (aa0)--(aa5)--(bb5)--(aa0);
\node at #1 [ddot]{};
\draw[int] (aa0)--($(aa0)!0.967!(aa5)$);
\draw[int] (aa0)--(bb5);
\node at ($#1+(#2+2.85*\legSpread:\labelDist*1.1)$)[]{{\footnotesize #3}};
\node at ($#1+(#2-2.85*\legSpread:\labelDist*1.1)$)[]{{\footnotesize #4}};
}

\newcommand{\legLabelled}[5]{\node at #1 [ddot]{};\draw[int] #1--($#1+(#2+\legSpread*6.5:\extLegLen)$);
\draw[int] #1--($#1+(#2-\legSpread*6.5:\extLegLen)$);\node at ($#1+(#2+\legSpread*6.5:\extLegLen*1.4)$)[]{{\footnotesize #3}};\node at ($#1+(#2-\legSpread*6.5:\extLegLen*1.4)$)[]{{\footnotesize #4}};\ifthenelse{#5=1}{\node[rotate=#2+90] at ($#1+(#2:\labelDist*0.8)$)[]{{\footnotesize $\cdots$}};}{};}

\newcommand{\tikzBox}[2][0.5]{\begin{tikzpicture}[scale=1,baseline=-2.05,rotate=0]\useasboundingbox ($\figScale*(-#1,-#1)$) rectangle ($\figScale*(#1,#1)$);#2\end{tikzpicture}}

\def\figScale{1.0}
\definecolor{optLegColour2}{rgb}{0.4,0.4,0.4}
\def\edgeLen{1*\figScale}

\pgfmathsetmacro{\pLen}{\edgeLen/(2*sin(72/2))}

\def\labelDist{\legLen*1.5}
\def\lineThickness{(1.25pt)}
\def\dotSize{(\figScale*12pt)}
\def\legSpread{4}
\def\dotSize{15*\figScale}
\def\extLegLen{1.12725*0.32*\figScale}
\def\labelDist{0.55*\figScale}
\def\ampSize{(1.1*\figScale*12pt)}

\def\legSpread{6}
\def\extLegLen{1.2725*0.34*\figScale}
\tikzset{int/.style={black,line width=\lineThickness,line cap=round,rounded corners=0.5pt}}
\tikzset{fullamp/.style={coordinate,minimum size=0.85*\ampSize,ball color=black!40,circle,text=white,inner sep=0}}
\tikzset{fullmhv/.style={coordinate,minimum size=0.85*\ampSize,ball color=mhvblue,circle,text=white,inner sep=0}}
\tikzset{fullmhvBar/.style={coordinate,minimum size=0.85*\ampSize,ball color=white,circle,text=white,inner sep=0}}
\tikzset{mhv/.style={fill=mhvblue,circle,draw=black,line width=\lineThickness,minimum size=0.375*\ampSize,text=white,inner sep=0}}
\tikzset{mhvBar/.style={fill=white,circle,draw=black,line width=\lineThickness,minimum size=0.375*\ampSize,text=white,inner sep=0}}
\tikzset{fullamp/.style={coordinate,minimum size=0.85*\ampSize,ball color=black!40,circle,text=white,inner sep=0}}
\tikzset{fullmhv/.style={coordinate,minimum size=0.85*\ampSize,ball color=mhvblue,circle,text=white,inner sep=0}}
\tikzset{fullmhvBar/.style={coordinate,minimum size=0.85*\ampSize,ball color=white,circle,text=white,inner sep=0}}
\tikzset{mhv/.style={fill=mhvblue,circle,draw=black,line width=\lineThickness,minimum size=0.375*\ampSize,text=white,inner sep=0}}
\tikzset{mhvBar/.style={fill=white,circle,draw=black,line width=\lineThickness,minimum size=0.375*\ampSize,text=white,inner sep=0}}

\tikzset{intTermCornerGen/.style={coordinate,minimum size=0.75*\ampSize,ball color=black!40,circle,text=white,inner sep=0}}
\tikzset{intTermCornerMHV/.style={coordinate,minimum size=0.75*\ampSize,ball color=mhvblue,circle,text=white,inner sep=0}}
\tikzset{intTermCornerMHVbar/.style={coordinate,minimum size=0.75*\ampSize,ball color=white,circle,text=white,inner sep=0}}

\tikzset{mhvAmp/.style={fill=mhvblue,circle,draw=black,line width=\lineThickness,minimum size=0.75*\ampSize,text=white,inner sep=0}}
\tikzset{mhvBarAmp/.style={fill=white,circle,draw=black,line width=\lineThickness,minimum size=0.75*\ampSize,text=white,inner sep=0}}
\tikzset{genAmp/.style={fill=black!30,circle,draw=black,line width=\lineThickness,minimum size=0.75*\ampSize,text=white,inner sep=0}}

\tikzset{mhvIntVertex/.style={fill=mhvblue,circle,draw=black,line width=\lineThickness,minimum size=0.375*\ampSize,text=white,inner sep=0}}
\tikzset{mhvBarIntVertex/.style={fill=white,circle,draw=black,line width=\lineThickness,minimum size=0.375*\ampSize,text=white,inner sep=0}}
\tikzset{ddot/.style={fill=black,circle,minimum size=0.35*\dotSize,inner sep=0}}

\definecolor{optLegColour}{rgb}{0.4,0.4,0.4}
\newcommand{\generalCorner}[2]{\coordinate (aa0) at #1;
\coordinate (aa5) at ($#1+(#2+\legSpread*3:\extLegLen)$);
\coordinate (bb5) at ($#1+(#2-\legSpread*3:\extLegLen)$);
\coordinate (aa1) at ($(aa0)!0.4!(aa5)$);\coordinate (aa2) at ($(aa0)!0.55!(aa5)$);\coordinate (aa3) at ($(aa0)!0.7!(aa5)$);\coordinate (aa4) at ($(aa0)!0.85!(aa5)$);
\coordinate (bb1) at ($(aa0)!0.4!(bb5)$);\coordinate (bb2) at ($(aa0)!0.55!(bb5)$);\coordinate (bb3) at ($(aa0)!0.7!(bb5)$);\coordinate (bb4) at ($(aa0)!0.85!(bb5)$);
\fill[optLegColour] (aa0)--(aa1)--(bb1)--(aa0);
\fill[optLegColour] (aa2)--(aa3)--(bb3)--(bb2)--(aa2);
\fill[optLegColour] (aa4)--(aa5)--(bb5)--(bb4)--(aa4);
\node at #1 [ddot]{};
\draw[int] (aa0)--(aa5);
}

\definecolor{optLegColour}{rgb}{0.5,0.5,0.5}
\definecolor{massiveCornerColour}{rgb}{0.5,0.5,0.5}

\newcommand{\generalCornerMassive}[2]{\coordinate (aa0) at #1;
\coordinate (aa5) at ($#1+(#2+\legSpread*3:\extLegLen)$);
\coordinate (bb5) at ($#1+(#2-\legSpread*3:\extLegLen)$);
\coordinate (aa1) at ($(aa0)!0.4!(aa5)$);\coordinate (aa2) at ($(aa0)!0.55!(aa5)$);\coordinate (aa3) at ($(aa0)!0.7!(aa5)$);\coordinate (aa4) at ($(aa0)!0.85!(aa5)$);
\coordinate (bb1) at ($(aa0)!0.4!(bb5)$);\coordinate (bb2) at ($(aa0)!0.55!(bb5)$);\coordinate (bb3) at ($(aa0)!0.7!(bb5)$);\coordinate (bb4) at ($(aa0)!0.85!(bb5)$);
\fill[massiveCornerColour] (aa0)--(aa5)--(bb5)--(aa0);
\node at #1 [ddot]{};
\draw[int] (aa0)--($(aa0)!0.967!(aa5)$);
\draw[int] (aa0)--(bb5);
}

\newcommand{\oneLoopChiralBoxInt}[9]{\def\rotn{-90}\def\edgeLen{1*\figScale}
\begin{tikzpicture}[scale=\figScale,baseline=-1.5]
\ifthenelse{\showTikzQ=1}{\useasboundingbox ($\figScale*(-0.7,-0.7)$) rectangle ($\figScale*(0.7,0.7)$);
\coordinate (v0) at (0,0);
\foreach\a in {0,...,4}{\coordinate (a\a) at ($(v0)+(\a*\rotn-90-\rotn/2:\edgeLen/2)$);};
\foreach\a[remember=\a as \la] in {0,...,4}{\ifthenelse{\a=0}{}{\draw[int](a\la)--(a\a)coordinate (e\a) at ($(a\la)!0.5!(a\a)$);}};
\ifthenelse{#1=3}{\generalCorner{(a0)}{-45};}{\ifthenelse{#1=2}{\generalCornerMassive{(a0)}{-45};}{\singleLeg{(a0)}{-45};}};
\ifthenelse{#2=3}{\generalCorner{(a1)}{225};}{\ifthenelse{#2=2}{\generalCornerMassive{(a1)}{225};}{\singleLeg{(a1)}{225};}};
\ifthenelse{#3=3}{\generalCorner{(a2)}{135};}{\ifthenelse{#3=2}{\generalCornerMassive{(a2)}{135};}{\singleLeg{(a2)}{135};}};
\ifthenelse{#4=3}{\generalCorner{(a3)}{45};}{\ifthenelse{#4=2}{\generalCornerMassive{(a3)}{45};}{\singleLeg{(a3)}{45};}};
\ifthenelse{#9=1}{\node[ddot] at (a0) {};\node[label={[label distance=2*\labelDist]-45:{#5}}] at (a0) {};\node[ddot] at (a1) {};\node[label={[label distance=2*\labelDist]225:{#6}}] at (a1) {};\node[ddot] at (a2) {};\ifthenelse{#3=1}{\node[label={[label distance=2*\labelDist]135:{#7}}] at (a2) {};}{\node[label={[label distance=4*\labelDist]177.5:{#7}}] at (a2) {};};\node[mhvBarIntVertex] at (a3) {};\node[label={[label distance=2*\labelDist]45:{#8}}] at (a3) {};}
{\ifthenelse{#9=2}{\node[ddot] at (a0) {};\node[label={[label distance=2*\labelDist]-45:{#5}}] at (a0) {};\node[ddot] at (a1) {};\ifthenelse{#2=1}{\node[label={[label distance=2*\labelDist]225:{#6}}] at (a1) {};}{\node[label={[label distance=4*\labelDist]255:{#6}}] at (a1) {};};\node[ddot] at (a2) {};\ifthenelse{#3=1}{\node[label={[label distance=2*\labelDist]135:{#7}}] at (a2) {};}{\node[label={[label distance=4*\labelDist]177.5:{#7}}] at (a2) {};};\node[mhvIntVertex] at (a3) {};\node[label={[label distance=2*\labelDist]45:{#8}}] at (a3) {};}
{\ifthenelse{#9=3}{\node[mhvBarIntVertex] at (a0) {};\node[label={[label distance=2*\labelDist]-45:{#5}}] at (a0) {};\node[ddot] at (a1) {};\node[label={[label distance=2*\labelDist]225:{#6}}] at (a1) {};\node[ddot] at (a2) {};\ifthenelse{#3=1}{\node[label={[label distance=2*\labelDist]135:{#7}}] at (a2) {};}{\node[label={[label distance=4*\labelDist]177.5:{#7}}] at (a2) {};};\node[mhvIntVertex] at (a3) {};\node[label={[label distance=2*\labelDist]45:{#8}}] at (a3) {};}
{\ifthenelse{#9=4}{\node[mhvIntVertex] at (a0) {};\node[label={[label distance=2*\labelDist]-45:{#5}}] at (a0) {};\node[ddot] at (a1) {};\node[label={[label distance=2*\labelDist]225:{#6}}] at (a1) {};\node[ddot] at (a2) {};\ifthenelse{#3=1}{\node[label={[label distance=2*\labelDist]135:{#7}}] at (a2) {};}{\node[xshift=-0.65*\figScale,yshift=-1.05*\figScale,label={[label distance=4.1*\labelDist]177.5:{#7}}] at (a2) {};};\node[mhvBarIntVertex] at (a3) {};\node[label={[label distance=2*\labelDist]45:{#8}}] at (a3) {};}
{\ifthenelse{#9=5}{\node[mhvIntVertex] at (a0) {};\node[label={[label distance=2*\labelDist]-45:{#5}}] at (a0) {};\node[mhvBarIntVertex] at (a1) {};\node[label={[label distance=2*\labelDist]225:{#6}}] at (a1) {};\node[ddot] at (a2) {};\ifthenelse{#3=1}{\node[label={[label distance=2*\labelDist]135:{#7}}] at (a2) {};}{\node[label={[label distance=4*\labelDist]177.5:{#7}}] at (a2) {};};\node[mhvBarIntVertex] at (a3) {};\node[label={[label distance=2*\labelDist]45:{#8}}] at (a3) {};}
{\ifthenelse{#9=6}{\node[ddot] at (a0) {};\node[label={[label distance=2*\labelDist]-45:{#5}}] at (a0) {};\node[mhvBarIntVertex] at (a1) {};\node[label={[label distance=2*\labelDist]201:{#6}}] at (a1) {};;\node[ddot] at (a2) {};\ifthenelse{#3=1}{\node[label={[label distance=2*\labelDist]135:{#7}}] at (a2) {};}{\node[xshift=-0.65*\figScale,yshift=-1.05*\figScale,label={[label distance=4.1*\labelDist]177.5:{#7}}] at (a2) {};};\node[mhvBarIntVertex] at (a3) {};\node[label={[label distance=2*\labelDist]45:{#8}}] at (a3) {};}
{\ifthenelse{#9=7}{\node[ddot] at (a0) {};\node[label={[label distance=2*\labelDist]-45:{#5}}] at (a0) {};\node[mhvIntVertex] at (a1) {};\node[label={[label distance=2*\labelDist]225:{#6}}] at (a1) {};\node[ddot] at (a2) {};\ifthenelse{#3=1}{\node[label={[label distance=2*\labelDist]135:{#7}}] at (a2) {};}{\node[label={[label distance=4*\labelDist]177.5:{#7}}] at (a2) {};};\node[mhvIntVertex] at (a3) {};\node[label={[label distance=2*\labelDist]45:{#8}}] at (a3) {};}{\ifthenelse{#9=8}{\node[mhvBarIntVertex] at (a0) {};\node[label={[label distance=2*\labelDist]-45:{#5}}] at (a0) {};\node[mhvIntVertex] at (a1) {};\node[label={[label distance=2*\labelDist]225:{#6}}] at (a1) {};\node[ddot] at (a2) {};\ifthenelse{#3=1}{\node[label={[label distance=2*\labelDist]135:{#7}}] at (a2) {};}{\node[label={[label distance=4*\labelDist]177.5:{#7}}] at (a2) {};};\node[mhvIntVertex] at (a3) {};\node[label={[label distance=2*\labelDist]45:{#8}}] at (a3) {};}{};};};};};};};};
}{\draw [line width=0.15] ($\figScale*(-0.6,-0.6)$) rectangle ($\figScale*(0.6,0.6)$);}
\end{tikzpicture}
}

\newcommand{\drawMerger}[2]{#1\hspace{5pt}\merge\hspace{5pt}#2}

\newcommand{\oneLoopBoxCoeff}[9]{\def\rotn{-90}\def\edgeLen{1*\figScale}
\begin{tikzpicture}[scale=\figScale,baseline=-1.5]
\ifthenelse{\showTikzQ=1}{\useasboundingbox ($\figScale*(-0.6,-0.6)$) rectangle ($\figScale*(0.6,0.6)$);
\coordinate (v0) at (0,0);
\foreach\a in {0,...,4}{\coordinate (a\a) at ($(v0)+(\a*\rotn-90-\rotn/2:\edgeLen/2)$);};
\foreach\a[remember=\a as \la] in {0,...,4}{\ifthenelse{\a=0}{}{\draw[int](a\la)--(a\a)coordinate (e\a) at ($(a\la)!0.5!(a\a)$);}};
\ifthenelse{#1=3}{\generalCorner{(a0)}{-45};}{\ifthenelse{#1=2}{\generalCornerMassive{(a0)}{-45};}{\singleLeg{(a0)}{-45};}};
\ifthenelse{#2=3}{\generalCorner{(a1)}{225};}{\ifthenelse{#2=2}{\generalCornerMassive{(a1)}{225};}{\singleLeg{(a1)}{225};}};
\ifthenelse{#3=3}{\generalCorner{(a2)}{135};}{\ifthenelse{#3=2}{\generalCornerMassive{(a2)}{135};}{\singleLeg{(a2)}{135};}};
\ifthenelse{#4=3}{\generalCorner{(a3)}{45};}{\ifthenelse{#4=2}{\generalCornerMassive{(a3)}{45};}{\singleLeg{(a3)}{45};}};
\ifthenelse{#9=1}{\node[genAmp] at (a0) {};\node[label={[label distance=2*\labelDist]-45:{#5}}] at (a0) {};\node[genAmp] at (a1) {};\node[xshift=9.65*\figScale,yshift=1.05*\figScale,label={[label distance=2*\labelDist]225:{#6}}] at (a1) {};\node[genAmp] at (a2) {};\ifthenelse{#3=1}{\node[label={[label distance=2*\labelDist]135:{#7}}] at (a2) {};}{\node[xshift=-0.65*\figScale,yshift=-1.05*\figScale,label={[label distance=4.1*\labelDist]177.5:{#7}}] at (a2) {};};\node[mhvBarAmp] at (a3) {};\node[label={[label distance=2*\labelDist]45:{#8}}] at (a3) {};}
{\ifthenelse{#9=2}{\node[genAmp] at (a0) {};\node[label={[label distance=2*\labelDist]-45:{#5}}] at (a0) {};\node[genAmp] at (a1) {};\ifthenelse{#2=1}{\node[label={[label distance=2*\labelDist]225:{#6}}] at (a1) {};}{\node[xshift=9.65*\figScale,yshift=1.05*\figScale,label={[label distance=2*\labelDist]225:{#6}}] at (a1) {};};\node[genAmp] at (a2) {};\ifthenelse{#3=1}{\node[label={[label distance=2*\labelDist]135:{#7}}] at (a2) {};}{\node[xshift=-0.65*\figScale,yshift=-1.05*\figScale,label={[label distance=4.1*\labelDist]177.5:{#7}}] at (a2) {};};\node[mhvAmp] at (a3) {};\node[label={[label distance=2*\labelDist]45:{#8}}] at (a3) {};}
{\ifthenelse{#9=3}{\node[mhvBarAmp] at (a0) {};\node[label={[label distance=2*\labelDist]-45:{#5}}] at (a0) {};\node[genAmp] at (a1) {};\node[label={[label distance=2*\labelDist]225:{#6}}] at (a1) {};\node[genAmp] at (a2) {};\ifthenelse{#3=1}{\node[label={[label distance=2*\labelDist]135:{#7}}] at (a2) {};}{\node[xshift=-0.65*\figScale,yshift=-1.05*\figScale,label={[label distance=4.1*\labelDist]177.5:{#7}}] at (a2) {};};\node[mhvAmp] at (a3) {};\node[label={[label distance=2*\labelDist]45:{#8}}] at (a3) {};}
{\ifthenelse{#9=4}{\node[mhvAmp] at (a0) {};\node[label={[label distance=2*\labelDist]-45:{#5}}] at (a0) {};\node[genAmp] at (a1) {};\node[label={[label distance=2*\labelDist]225:{#6}}] at (a1) {};\node[genAmp] at (a2) {};\ifthenelse{#3=1}{\node[label={[label distance=2*\labelDist]135:{#7}}] at (a2) {};}{\node[xshift=-0.65*\figScale,yshift=-1.05*\figScale,label={[label distance=4.1*\labelDist]177.5:{#7}}] at (a2) {};};\node[mhvBarAmp] at (a3) {};\node[label={[label distance=2*\labelDist]45:{#8}}] at (a3) {};}
{\ifthenelse{#9=5}{\node[mhvAmp] at (a0) {};\node[label={[label distance=2*\labelDist]-45:{#5}}] at (a0) {};\node[mhvBarAmp] at (a1) {};\node[label={[label distance=2*\labelDist]225:{#6}}] at (a1) {};\node[genAmp] at (a2) {};\ifthenelse{#3=1}{\node[label={[label distance=2*\labelDist]135:{#7}}] at (a2) {};}{\node[label={[label distance=4*\labelDist]177.5:{#7}}] at (a2) {};};\node[mhvBarAmp] at (a3) {};\node[label={[label distance=2*\labelDist]45:{#8}}] at (a3) {};}
{\ifthenelse{#9=6}{\node[genAmp] at (a0) {};\node[label={[label distance=2*\labelDist]-45:{#5}}] at (a0) {};\node[mhvBarAmp] at (a1) {};\node[label={[label distance=2*\labelDist]205:{#6}}] at (a1) {};\node[genAmp] at (a2) {};\ifthenelse{#3=1}{\node[label={[label distance=2*\labelDist]135:{#7}}] at (a2) {};}{\node[xshift=-0.65*\figScale,yshift=-1.05*\figScale,label={[label distance=4.1*\labelDist]177.5:{#7}}] at (a2) {};};\node[mhvBarAmp] at (a3) {};\node[label={[label distance=2*\labelDist]45:{#8}}] at (a3) {};}{\ifthenelse{#9=7}{\node[genAmp] at (a0) {};\node[label={[label distance=2*\labelDist]-45:{#5}}] at (a0) {};\node[mhvAmp] at (a1) {};\node[label={[label distance=2*\labelDist]205:{#6}}] at (a1) {};\node[genAmp] at (a2) {};\ifthenelse{#3=1}{\node[label={[label distance=2*\labelDist]135:{#7}}] at (a2) {};}{\node[xshift=-0.65*\figScale,yshift=-1.05*\figScale,label={[label distance=4.1*\labelDist]177.5:{#7}}] at (a2) {};};\node[mhvAmp] at (a3) {};\node[label={[label distance=2*\labelDist]45:{#8}}] at (a3) {};}{\ifthenelse{#9=8}{\node[mhvBarAmp] at (a0) {};\node[label={[label distance=2*\labelDist]-45:{#5}}] at (a0) {};\node[mhvAmp] at (a1) {};\node[label={[label distance=2*\labelDist]225:{#6}}] at (a1) {};\node[genAmp] at (a2) {};\ifthenelse{#3=1}{\node[label={[label distance=2*\labelDist]135:{#7}}] at (a2) {};}{\node[label={[label distance=4*\labelDist]177.5:{#7}}] at (a2) {};};\node[mhvAmp] at (a3) {};\node[label={[label distance=2*\labelDist]45:{#8}}] at (a3) {};}{};};};};};};};};
}{\draw [line width=0.15] ($\figScale*(-0.6,-0.6)$) rectangle ($\figScale*(0.6,0.6)$);}
\end{tikzpicture}
}

\newcommand{\oneLoopLSxInt}[9]{\def\rotn{-90}\def\edgeLen{1*\figScale}
\begin{tikzpicture}[scale=\figScale,baseline=-1.5]
\ifthenelse{\showTikzQ=1}{\useasboundingbox ($\figScale*(-0.7,-0.7)$) rectangle ($\figScale*(0.7,0.7)$);
\coordinate (v0) at (0,0);
\foreach\a in {0,...,4}{\coordinate (a\a) at ($(v0)+(\a*\rotn-90-\rotn/2:\edgeLen/2)$);};
\foreach\a[remember=\a as \la] in {0,...,4}{\ifthenelse{\a=0}{}{\draw[int](a\la)--(a\a)coordinate (e\a) at ($(a\la)!0.5!(a\a)$);}};
\ifthenelse{#1=3}{\generalCorner{(a0)}{-45};}{\ifthenelse{#1=2}{\generalCornerMassive{(a0)}{-45};}{\singleLeg{(a0)}{-45};}};
\ifthenelse{#2=3}{\generalCorner{(a1)}{225};}{\ifthenelse{#2=2}{\generalCornerMassive{(a1)}{225};}{\singleLeg{(a1)}{225};}};
\ifthenelse{#3=3}{\generalCorner{(a2)}{135};}{\ifthenelse{#3=2}{\generalCornerMassive{(a2)}{135};}{\singleLeg{(a2)}{135};}};
\ifthenelse{#4=3}{\generalCorner{(a3)}{45};}{\ifthenelse{#4=2}{\generalCornerMassive{(a3)}{45};}{\singleLeg{(a3)}{45};}};
\ifthenelse{#9=10}{\node[intTermCornerMHV] at (a0) {};\node[intTermCornerMHVbar] at (a3) {};\node[intTermCornerMHVbar] at (a1) {};\node[intTermCornerMHV] at (a2) {};\node[label={[label distance=2*\labelDist]+45:{#8}}] at (a3) {};};
\ifthenelse{#9=11}{\node[intTermCornerMHV] at (a0) {};\node[intTermCornerMHVbar] at (a3) {};\node[intTermCornerMHVbar] at (a1) {};\node[intTermCornerMHV] at (a2) {};\node[label={[label distance=2*\labelDist]-45:{#5}}] at (a0) {};};
\ifthenelse{#9=1}{\node[intTermCornerGen] at (a0) {};\node[label={[label distance=2*\labelDist]-45:{#5}}] at (a0) {};\node[intTermCornerGen] at (a1) {};\node[xshift=9.65*\figScale,yshift=1.05*\figScale,label={[label distance=2*\labelDist]225:{#6}}] at (a1) {};\node[intTermCornerGen] at (a2) {};\ifthenelse{#3=1}{\node[label={[label distance=2*\labelDist]135:{#7}}] at (a2) {};}{\node[xshift=-0.65*\figScale,yshift=-1.05*\figScale,label={[label distance=4.1*\labelDist]177.5:{#7}}] at (a2) {};};\node[intTermCornerMHVbar] at (a3) {};\node[label={[label distance=2*\labelDist]45:{#8}}] at (a3) {};}
{\ifthenelse{#9=2}{\node[intTermCornerGen] at (a0) {};\node[label={[label distance=2*\labelDist]-45:{#5}}] at (a0) {};\node[intTermCornerGen] at (a1) {};\ifthenelse{#2=1}{\node[label={[label distance=2*\labelDist]225:{#6}}] at (a1) {};}{\node[xshift=9.65*\figScale,yshift=1.05*\figScale,label={[label distance=2*\labelDist]225:{#6}}] at (a1) {};};\node[intTermCornerGen] at (a2) {};\ifthenelse{#3=1}{\node[label={[label distance=2*\labelDist]135:{#7}}] at (a2) {};}{\node[xshift=-0.65*\figScale,yshift=-1.05*\figScale,label={[label distance=4.1*\labelDist]177.5:{#7}}] at (a2) {};};\node[intTermCornerMHV] at (a3) {};\node[label={[label distance=2*\labelDist]45:{#8}}] at (a3) {};}
{\ifthenelse{#9=3}{\node[intTermCornerMHVbar] at (a0) {};\node[label={[label distance=2*\labelDist]-45:{#5}}] at (a0) {};\node[intTermCornerGen] at (a1) {};\node[label={[label distance=2*\labelDist]225:{#6}}] at (a1) {};\node[intTermCornerGen] at (a2) {};\ifthenelse{#3=1}{\node[label={[label distance=2*\labelDist]135:{#7}}] at (a2) {};}{\node[xshift=-0.65*\figScale,yshift=-1.05*\figScale,label={[label distance=4.1*\labelDist]177.5:{#7}}] at (a2) {};};\node[intTermCornerMHV] at (a3) {};\node[label={[label distance=2*\labelDist]45:{#8}}] at (a3) {};}
{\ifthenelse{#9=4}{\node[intTermCornerMHV] at (a0) {};\node[label={[label distance=2*\labelDist]-45:{#5}}] at (a0) {};\node[intTermCornerGen] at (a1) {};\node[label={[label distance=2*\labelDist]225:{#6}}] at (a1) {};\node[intTermCornerGen] at (a2) {};\ifthenelse{#3=1}{\node[label={[label distance=2*\labelDist]135:{#7}}] at (a2) {};}{\node[xshift=-0.65*\figScale,yshift=-1.05*\figScale,label={[label distance=4.1*\labelDist]177.5:{#7}}] at (a2) {};};\node[intTermCornerMHVbar] at (a3) {};\node[label={[label distance=2*\labelDist]45:{#8}}] at (a3) {};}
{\ifthenelse{#9=5}{\node[intTermCornerMHV] at (a0) {};\node[label={[label distance=2*\labelDist]-45:{#5}}] at (a0) {};\node[intTermCornerMHVbar] at (a1) {};\node[label={[label distance=2*\labelDist]225:{#6}}] at (a1) {};\node[intTermCornerGen] at (a2) {};\ifthenelse{#3=1}{\node[label={[label distance=2*\labelDist]135:{#7}}] at (a2) {};}{\node[label={[label distance=4*\labelDist]177.5:{#7}}] at (a2) {};};\node[intTermCornerMHVbar] at (a3) {};\node[label={[label distance=2*\labelDist]45:{#8}}] at (a3) {};}
{\ifthenelse{#9=6}{\node[intTermCornerGen] at (a0) {};\node[label={[label distance=2*\labelDist]-45:{#5}}] at (a0) {};\node[intTermCornerMHVbar] at (a1) {};\node[label={[label distance=2*\labelDist]205:{#6}}] at (a1) {};\node[intTermCornerGen] at (a2) {};\ifthenelse{#3=1}{\node[label={[label distance=2*\labelDist]135:{#7}}] at (a2) {};}{\node[xshift=-0.65*\figScale,yshift=-1.05*\figScale,label={[label distance=4.1*\labelDist]177.5:{#7}}] at (a2) {};};\node[intTermCornerMHVbar] at (a3) {};\node[label={[label distance=2*\labelDist]45:{#8}}] at (a3) {};}{\ifthenelse{#9=7}{\node[intTermCornerGen] at (a0) {};\node[label={[label distance=2*\labelDist]-45:{#5}}] at (a0) {};\node[intTermCornerMHV] at (a1) {};\node[label={[label distance=2*\labelDist]205:{#6}}] at (a1) {};\node[intTermCornerGen] at (a2) {};\ifthenelse{#3=1}{\node[label={[label distance=2*\labelDist]135:{#7}}] at (a2) {};}{\node[xshift=-0.65*\figScale,yshift=-1.05*\figScale,label={[label distance=4.1*\labelDist]177.5:{#7}}] at (a2) {};};\node[intTermCornerMHV] at (a3) {};\node[label={[label distance=2*\labelDist]45:{#8}}] at (a3) {};}{\ifthenelse{#9=8}{\node[intTermCornerMHVbar] at (a0) {};\node[label={[label distance=2*\labelDist]-45:{#5}}] at (a0) {};\node[intTermCornerMHV] at (a1) {};\node[label={[label distance=2*\labelDist]225:{#6}}] at (a1) {};\node[intTermCornerGen] at (a2) {};\ifthenelse{#3=1}{\node[label={[label distance=2*\labelDist]135:{#7}}] at (a2) {};}{\node[label={[label distance=4*\labelDist]177.5:{#7}}] at (a2) {};};\node[intTermCornerMHV] at (a3) {};\node[label={[label distance=2*\labelDist]45:{#8}}] at (a3) {};}{};};};};};};};};
}{\draw [line width=0.15] ($\figScale*(-0.6,-0.6)$) rectangle ($\figScale*(0.6,0.6)$);}
\end{tikzpicture}
}

\newcommand{\oneLoopMHVInt}[8]{\def\rotn{-90}\def\edgeLen{1*\figScale}
\begin{tikzpicture}[scale=\figScale,baseline=-1.5]
\ifthenelse{\showTikzQ=1}{\useasboundingbox ($\figScale*(-0.7,-0.7)$) rectangle ($\figScale*(0.7,0.7)$);
\coordinate (v0) at (0,0);
\foreach\a in {0,...,4}{\coordinate (a\a) at ($(v0)+(\a*\rotn-90-\rotn/2:\edgeLen/2)$);};
\foreach\a[remember=\a as \la] in {0,...,4}{\ifthenelse{\a=0}{}{\draw[int](a\la)--(a\a)coordinate (e\a) at ($(a\la)!0.5!(a\a)$);}};
\ifthenelse{#1=3}{\generalCorner{(a0)}{-45};}{\ifthenelse{#1=2}{\generalCornerMassive{(a0)}{-45};}{\singleLeg{(a0)}{-45};}};
\ifthenelse{#2=3}{\generalCorner{(a1)}{225};}{\ifthenelse{#2=2}{\generalCornerMassive{(a1)}{225};}{\singleLeg{(a1)}{225};}};
\ifthenelse{#3=3}{\generalCorner{(a2)}{135};}{\ifthenelse{#3=2}{\generalCornerMassive{(a2)}{135};}{\singleLeg{(a2)}{135};}};
\ifthenelse{#4=3}{\generalCorner{(a3)}{45};}{\ifthenelse{#4=2}{\generalCornerMassive{(a3)}{45};}{\singleLeg{(a3)}{45};}};
\node[intTermCornerMHV] at (a0) {};\node[label={[label distance=2*\labelDist]-45:{#5}}] at (a0) {};\node[intTermCornerMHVbar] at (a1) {};\node[label={[label distance=2*\labelDist]205:{#6}}] at (a1) {};\node[intTermCornerMHV] at (a2) {};\ifthenelse{#3=1}{\node[label={[label distance=2*\labelDist]135:{#7}}] at (a2) {};}{\node[label={[label distance=4*\labelDist]177.5:{#7}}] at (a2) {};};\node[intTermCornerMHVbar] at (a3) {};\node[label={[label distance=2*\labelDist]45:{#8}}] at (a3) {};
}{\draw [line width=0.15] ($\figScale*(-0.6,-0.6)$) rectangle ($\figScale*(0.6,0.6)$);}
\end{tikzpicture}
}

\newcommand{\oneLoopScalarBox}[4]{\def\rotn{-90}\def\edgeLen{1*\figScale}
\begin{tikzpicture}[scale=\figScale,baseline=-1.00]
\ifthenelse{\showTikzQ=1}{\useasboundingbox ($\figScale*(-0.6,-0.6)$) rectangle ($\figScale*(0.6,0.6)$);
\coordinate (v0) at (0,0);
\foreach\a in {0,...,4}{\coordinate (a\a) at ($(v0)+(\a*\rotn-90-\rotn/2:\edgeLen/2)$);};
\foreach\a[remember=\a as \la] in {0,...,4}{\ifthenelse{\a=0}{}{\draw[int](a\la)--(a\a)coordinate (e\a) at ($(a\la)!0.5!(a\a)$);}};
\ifthenelse{#1=3}{\generalCorner{(a0)}{-45};}{\ifthenelse{#1=2}{\generalCornerMassive{(a0)}{-45};}{\singleLeg{(a0)}{-45};}};
\ifthenelse{#2=3}{\generalCorner{(a1)}{225};}{\ifthenelse{#2=2}{\generalCornerMassive{(a1)}{225};}{\singleLeg{(a1)}{225};}};
\ifthenelse{#3=3}{\generalCorner{(a2)}{135};}{\ifthenelse{#3=2}{\generalCornerMassive{(a2)}{135};}{\singleLeg{(a2)}{135};}};
\ifthenelse{#4=3}{\generalCorner{(a3)}{45};}{\ifthenelse{#4=2}{\generalCornerMassive{(a3)}{45};}{\singleLeg{(a3)}{45};}};
}{\draw [line width=0.15] ($\figScale*(-0.6,-0.6)$) rectangle ($\figScale*(0.6,0.6)$);}
\end{tikzpicture}
}

\let\olditemize\itemize\renewcommand{\itemize}{\vspace{-2pt}\olditemize\setlength{\itemsep}{1pt}\setlength{\parskip}{0pt}\setlength{\parsep}{-0pt}}
\let\oldenumerate\enumerate\renewcommand{\enumerate}{\vspace{-4pt}\oldenumerate\setlength{\itemsep}{1pt}\setlength{\parskip}{0pt}\setlength{\parsep}{0pt}}
\makeatletter\renewcommand\section{\addtocontents{toc}{\protect\addvspace{-2.25\p@}}\@startsection {section}{1}{\z@}{-0.0ex \@plus .2ex \@minus 0.2ex}{1ex \@plus.1ex\@minus .5ex}{\normalfont\large\bfseries}}
\renewcommand\subsection{\addtocontents{toc}{\protect\addvspace{-2.5\p@}}\@startsection {subsection}{1}{\z@}{0.5ex \@plus .2ex \@minus 0.2ex}{0.75ex \@plus.1ex\@minus .5ex}{\normalfont\bfseries}}

\def\showTikzQ{0}

\definecolor{mhvBlue}{rgb}{0.3,0.2,0.75}
\definecolor{fRed}{rgb}{0.48,0.02824,0.18824}
\definecolor{cut2}{rgb}{0.18824,0.18824,0.48}
\definecolor{cut1}{rgb}{0.48,0.02824,0.18824}
\definecolor{hblue}{rgb}{0,0,0.575}
\definecolor{hred}{rgb}{0.475,0.0,0.15}
\definecolor{dred}{rgb}{0.575,0.4,0.45}

\DeclareMathOperator*{\Res}{\mathrm{Res}}

\newcommand{\eq}[1]{\vspace{-0.5pt}\begin{equation}#1\vspace{-0.5pt}\end{equation}}

\newcommand{\fwbox}[2]{\text{\makebox[#1][c]{$\hspace{-150pt}\displaystyle#2\hspace{-150pt}$}}}
\newcommand{\fwboxL}[2]{\text{\makebox[#1][l]{$#2$}}}
\newcommand{\fwboxR}[2]{\text{\makebox[#1][r]{$#2$}}}
\newcommand{\equivR}{\fwbox{14.5pt}{\hspace{-0pt}\fwboxR{0pt}{\raisebox{0.47pt}{\hspace{1.25pt}:\hspace{-4pt}}}=\fwboxL{0pt}{}}}
\newcommand{\equivL}{\fwbox{14.5pt}{\fwboxR{0pt}{}=\fwboxL{0pt}{\raisebox{0.47pt}{\hspace{-4pt}:\hspace{1.25pt}}}}}
\newcommand{\fig}[3]{\raisebox{#1}{\includegraphics[scale=#2]{#3}}}

\newcommand{\bigger}[1]{\raisebox{-0.95pt}{\scalebox{1.25}{$#1$}}}
\newcommand{\mi}{\raisebox{0.75pt}{\scalebox{0.75}{$\hspace{-0.5pt}\,-\,\hspace{-0.5pt}$}}}
\newcommand{\pl}{\raisebox{0.75pt}{\scalebox{0.75}{$\hspace{-0.5pt}\,+\,\hspace{-0.5pt}$}}}
\renewcommand{\phi}{\varphi}

\renewcommand{\tilde}{\widetilde}
\newcommand{\ab}[1]{\langle #1\rangle}
\renewcommand{\sb}[1]{[ #1]}

\newcommand{\x}[2]{{\color{black}(}\hspace{-0.85pt}{\color{black}#1}\hspace{-0.25pt}{\color{black}|}\hspace{-0.25pt}{\color{black}#2}\hspace{-0.85pt}{\color{black})}}

\definecolor{hblue}{rgb}{0,0,0.575}
\definecolor{hred}{rgb}{0.575,0.0,0.225}
\definecolor{totalCount}{rgb}{0,0,0.575}
\definecolor{topCount}{rgb}{0.575,0.0,0.225}
\definecolor{dim}{rgb}{0.55,0.55,0.55}
\definecolor{deemph}{rgb}{0.25,0.25,0.25}

\graphicspath{{figures/}}
\def\showTikzQ{1}

\title{\texorpdfstring{~\\[-20pt]{\huge \mbox{{\emph{Locally}}-Finite Observables in sYM}}\\[-6pt]}{Locally Finite Observables in sYM}}
\author[\ast,\dagger]{\vspace{-26pt}Jacob~L.~Bourjaily,}\emailAdd{bourjaily@psu.edu}
\author[\ast]{Cameron~Langer,}\emailAdd{ckl5552@psu.edu}
\author[\ast]{Kokkimidis~Patatoukos}\emailAdd{kzp326@psu.edu}
\affiliation[\ast]{Institute for Gravitation and the Cosmos, Department of Physics,\\Pennsylvania State University, University Park, PA 16802, USA}
\affiliation[\dagger]{Niels Bohr International Academy and Discovery Center, Niels Bohr Institute,\\University of Copenhagen, Blegdamsvej 17, DK-2100, Copenhagen \O, Denmark}
\abstract{%
A \emph{locally-finite} observable is one for which there is no region of divergence anywhere in the space of real loop momenta; it can therefore be computed (in principle) without regularization. In this work, we prove that \emph{all} two-loop ratio functions in planar, maximally supersymmetric Yang-Mills theory are locally-finite.}
\preprint{}

\begin{document}
\maketitle
\pagenumbering{roman}
%
\setcounter{page}{1}\vspace{0pt}\setcounter{section}{0}%
\pagenumbering{arabic}
\vspace{0pt}\section{Introduction and Overview}\label{sec:introduction}\vspace{-0pt}
Many of the recent advances in our understanding of and ability to compute scattering amplitudes in quantum field theory have relied on the simple---yet surprisingly powerful---idea of separating the two problems of constructing the loop integrand from carrying out loop integration. Indeed, in the case of the planar limit of maximally supersymmetric ($\mathcal{N}\!=\!4$) Yang-Mills theory (sYM), {the} loop integrand is a perfectly well-defined rational function which can be constructed from knowledge of its residues. More broadly, the systematic development of computational tools to determine the all-loop integrand has been a continuous source of insight into the unanticipated simplicity of an increasingly large class of field theories. In particular, investigating the structure of the loop integrand in planar sYM led to the discoveries of dual-conformal symmetry \cite{Drummond:2006rz, Alday:2007hr, Drummond:2008vq}, tree- and loop-level recursion relations \cite{BCF,BCFW,ArkaniHamed:2010kv,Benincasa:2015zna}, connections to Grassmannian geometry \cite{ArkaniHamed:2009dn,ArkaniHamed:2009dg,ArkaniHamed:2012nw,Arkani-Hamed:2013jha}, and the development of generalized \cite{Bern:1993kr,Bern:1994cg, Bern:1994zx,Britto:2004nc,Bern:2006ew,Anastasiou:2006jv,Bern:2007ct,Cachazo:2008vp,Bern:2008ap,Berger:2008sj,Abreu:2017xsl} and prescriptive \cite{ArkaniHamed:2010gh,Bourjaily:2013mma,Bourjaily:2015jna,Bourjaily:2017wjl,Bourjaily:2018omh,Bourjaily:2019iqr,Bourjaily:2019gqu,Bourjaily:2020qca} unitarity. 

Despite the significant advances made in representing loop integrands, the problem of loop \emph{integration} remains an exceedingly difficult one. In particular, the issues of regularization and renormalization seem unavoidable intermediate requirements of the computation of most observable quantities. Particularly relevant to the case of planar sYM is the fact that the scattering amplitudes themselves (as for any theory with massless particles) are infrared (IR) divergent and require regularization. While it is not \emph{a priori} clear how much of the simplicity of loop integrands remains post-integration, there exists an impressive body of evidence that certain infrared-safe quantities do in fact preserve many integrand-level properties (such as dual-conformal invariance). These include the remainder function \cite{Anastasiou:2003kj,Bern:2005iz,Bern:2008ap,DelDuca:2010zg,Goncharov:2010jf}, defined as the ratio of the maximally-helicity-violating (MHV) amplitude and the Bern-Dixon-Smirnov (BDS) ansatz, as well as the ratio function \cite{Drummond:2008vq}, which is the ratio of the $\text{N}^k$MHV amplitude to the MHV amplitude.  

In practice, computing finite observables such as the ratio function has required the explicit cancellation of infrared divergences among different loop orders. However, the difficulties of direct integration of multi-loop integrands has thus far prevented this approach from seriously competing with bootstrap methods \cite{Dixon:2011nj,Dixon:2013eka, Dixon:2014voa,Dixon:2014iba,Dixon:2015iva,Dixon:2016apl,Caron-Huot:2016owq,Caron-Huot:2019vjl,Caron-Huot:2019bsq,Drummond:2014ffa,Dixon:2016nkn,Dixon:2014xca}, for example, which sidestep both the loop integrand and integration issues entirely and have been used to determine the remainder and ratio functions to impressively high loop orders. Despite these remarkable achievements, these methods are limited in their applicability; in particular, they have little to say regarding transcendental functions which are known \cite{Bourjaily:2017bsb,Bourjaily:2018ycu,Bourjaily:2018yfy,Bourjaily:2020hjv} to be required at two loops and beyond (for large enough multiplicities). 

Although the ratio function is \emph{defined} at a given loop order as a combination of divergent amplitudes, one may ask if it is possible to do better at the integrand-level. Namely, can the ratio function be represented in terms of individually, infrared-finite integrals, eliminating the need for any regularization? At one loop, at least, the answer is a positive one: the basis described in \cite{ArkaniHamed:2010gh} and the work of \cite{Bourjaily:2013mma} demonstrate that all infrared finite observables in sYM can be represented directly in terms of individually, locally finite integrals. In this work, we provide a non-constructive affirmative answer to this question at two loops by demonstrating the cancellation of all infrared divergent regions \emph{at the integrand-level}. Along the way, we use the correspondence between collinear and soft regions of loop momentum space and infrared divergences to introduce the notion of the `local finiteness' of a two-loop integral, which can be diagnosed by simple tests at the integrand-level. This classification of infrared behavior should prove useful in the construction of manifestly finite integrals, which have already proven to be invaluable in a variety of contexts \cite{Brandhuber:2008pf,Drummond:2008bq,Drummond:2008cr,ArkaniHamed:2009dn,Mason:2009qx,ArkaniHamed:2009vw,Korchemsky:2010ut,ArkaniHamed:2010gh,Bourjaily:2015jna,Bourjaily:2019iqr,Bourjaily:2019gqu,Herrmann:2020oud}.

\vspace{0pt}%
\subsection{Organization and Outline}\label{subsec:outline}\vspace{0pt}
In this work we  examine the infrared behavior of the two-loop ratio function, at the integrand-level in planar, maximally supersymmetric ($\mathcal{N}\!=\!4$) Yang-Mills theory (sYM) and show that it (the ratio function) is locally finite. Starting from the local representations of all planar $\mathcal{N}\!=\!4$ two-loop integrands as introduced in \cite{Bourjaily:2015jna}, we demonstrate how the divergences associated with particular collinear and soft-collinear regions of loop momentum space cancel at the integrand-level between the $\text{N}^k$MHV$\times$MHV and MHV$\times$MHV terms in the ratio function. The two kinds of cancellations which arise can be understood either as grouping of terms into BCFW recursions of the sets, or as residue theorems on particular sub-leading singularities. 

This work is organized as follows. In section~\ref{sec:local_vs_spurious_finiteness} we introduce the notion of the `local' finiteness of a loop integrand, and contrast this with the weaker notion of being \emph{merely} (or `spuriously') finite in some particular regularization scheme. We call integrands `spuriously' finite if their finiteness is regularization-scheme-dependent. In section \ref{sec:local_vs_spurious_finiteness} we classify all possible locally-divergent regions of loop momentum space responsible for infrared divergences at two loops, and enumerate every potentially-divergent merge-generated integrand relevant for the ratio function in planar sYM. In section \ref{sec:main_result}, after reviewing the less-divergent representation of the two-loop integrand in \cite{Bourjaily:2015jna} we use the classification of divergences to demonstrate the cancellation of every local divergence in the two-loop ratio function, thus demonstrating its local finiteness. Finally, we summarize the main implications of this work and provide some potentially interesting avenues for future research.

\vspace{-0pt}\section{{\bf\emph{Actual}} (`Local') Finiteness vs.\ `\emph{Spurious}' Finiteness}\label{sec:local_vs_spurious_finiteness}\vspace{-0pt}
%

\vspace{-0pt}\subsection{Examples of Spurious Finiteness at One and Two Loops}\label{subsec:exempli_gratia}\vspace{-0pt}
%
%
The infrared singularities of loop amplitudes involving massless particles can be diagnosed by examining the behavior of integrands in the \emph{collinear} region: where a loop momentum  becomes proportional to a massless external particle's momentum, $\ell\propto p_a$, with $p_a^2{=}0$. This co-dimension three configuration corresponds to a composite residue and is associated with the following on-shell function:
\vspace{-10pt}\eq{\fig{-22.5pt}{1}{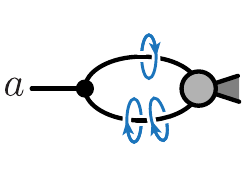}\;.\vspace{-10pt}\label{collinear_OS}}
The collinear region has a simple description in the \emph{dual momentum} co-ordinates $x_a$, which trivialize momentum conservation through the definition $p_a\equivL x_{a+1}{-}x_a$. The inverse propagators are given in dual co-ordinates as
\eq{\x{a}{b}\equivR(x_a-x_b)^2=(p_a+\ldots+p_{b-1})^2,\quad \x{\ell_i}{a}\equivR(x_{\ell_i}-x_a)^2,\quad \x{\ell_i}{\ell_j}\equivR(x_{\ell_i}-x_{\ell_j})^2,}
where the $x_{\ell_i}$ are the dual points associated to the loop momenta $\ell_i$. In these variables, the collinear region of loop momentum space (at one loop) associated with the on-shell function (\ref{collinear_OS}) can be approached as follows. First, one cuts either $\x{\ell}{a}=0$ or $\x{\ell}{a{+}1}=0$, upon which the other propagator factorizes into two pieces; cutting both of these factors (at co-dimension two) corresponds to the collinear residue. To be slightly more explicit, if we identify the loop momentum flowing into the leg $a$ as $\ell$, then on the support of $\ell^2=0$ we can write $\ell\equivL\lambda_\ell\tilde{\lambda}_\ell$ in terms of spinor helicity variables; in terms of these, the second inverse propagator becomes $(\ell{+}a)^2\rightarrow\ab{\ell\,a}\sb{\ell\,a}$. The collinear region is where \emph{both} of these factors vanish, which sets $\ell\propto p_a$. In this work, it will be useful to express this condition in terms of dual momenta, where the aforementioned collinear configuration associated with leg $a$ can be parametrized as
\eq{\label{eq:coll_description} \ell=\alpha\, x_a+(1-\alpha)\,x_{a+1}\,,}
where $\alpha$ is the single remaining degree of freedom in the loop momentum on this triple-cut. Any one-loop integral whose integrand has support on such a co-dimension three residue inescapably requires regularization to evaluate, and will generically yield \emph{at least} $1/\epsilon$, $\log(m^2)$ \cite{Alday:2009zm} or $\log(\delta)$ divergences in the dimensional-regularization (see \mbox{e.g.~\cite{Bollini:1972ui,tHooft:1972tcz,Bern:1992em}}), mass-(or `Higgs'-)regularization (see \mbox{e.g.~\cite{Alday:2009zm}}) and conformal-regularization \cite{Bourjaily:2013mma} schemes, respectively.

The so-called `soft-collinear' regions correspond to co-dimension four residues where the loop momentum is collinear to consecutive massless momenta (or, equivalently, the dual loop momentum is \emph{coincident} with a dual momentum coordinate), and corresponds to the following on-shell function:
\eq{\fig{-25pt}{1}{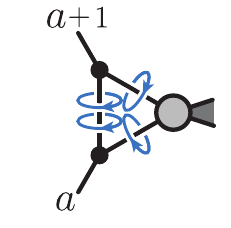}\;.}
In the parametrization of (\ref{eq:coll_description}), the soft-collinear singularities where $\ell=x_a$ or $\ell=x_{a{+}1}$ correspond to poles at the locations $\alpha\rightarrow1$ and $\alpha\rightarrow0$, respectively. An integrand with support on a soft-collinear residue will generically yield $1/\epsilon^2$, $\log^2(m^2)$ or $\log^2(\delta)$ divergences upon integration in the dimensional-regularization, mass-regularization and dual-conformal-regularization schemes, respectively. To be clear, throughout this work, the statement that an integral is `$\log^k(\delta)$-divergent' should be understood to refer to the \emph{highest} degree of divergence present in the integrated result. Of course, generically $\log^k(\delta)$-divergent integrals will also have sub-leading divergences. 

Because the one-loop collinear region is parity-invariant, this yields a simple criterion to test the convergence of a one-loop integral: multiply the integrand by two adjacent propagators, and check if the result vanishes when evaluated in the collinear region (\ref{eq:coll_description}). In accordance with \cite{Bourjaily:2013mma}, we define a \emph{truly}-convergent or `locally finite' one-loop integral to be one whose integrand vanishes in every collinear region. Constructing locally finite one-loop integrands is especially simple using the `chiral' numerators described at length in e.g. \cite{Bourjaily:2013mma,Bourjaily:2019iqr}. In fact, the `chiral octagons' introduced in \cite{ArkaniHamed:2010gh} can be used to make the infrared structure of any one-loop object completely manifest, as they naturally separate into manifestly locally finite and divergent terms---and no (non-vanishing) combination of divergent integrands is convergent. 

Our definition of local finiteness is to be contrasted with the weaker requirement of being \emph{merely} (i.e.\ `spuriously') finite in a particular regularization scheme. The existence of spuriously-finite integrals at one-loop have been well-studied in the literature, where the existence of dual-conformal symmetry was used as a proxy for local (actual, scheme-independent) finiteness (see \mbox{e.g.~\cite{Drummond:2008bq,Brandhuber:2009xz,Elvang:2009ya}} for examples of combinations of one-loop box integrals that are spuriously finite in dimensional regularization). A spuriously finite combination of one loop integrals is easier to find for the conformal regulator; for example, the following sum of four box integrals:
\eq{\label{spurious_boxes}\def\rotn{-90}\def\edgeLen{1*\figScale}\mathcal{I}_{\text{spurious}}\equivR\begin{tikzpicture}[scale=\figScale,baseline=-1.00]
\coordinate (v0) at (0,0);
\foreach\a in {0,...,4}{\coordinate (a\a) at ($(v0)+(\a*\rotn-90-\rotn/2:\edgeLen/2)$);};
\foreach\a[remember=\a as \la] in {0,...,4}{\ifthenelse{\a=0}{}{\draw[int](a\la)--(a\a)coordinate (e\a) at ($(a\la)!0.5!(a\a)$);}};
\singleLegLabelled{(a0)}{-45}{$2$};
\generalCornerMassiveLabelled{(a1)}{225}{$3$}{$6$};
\singleLegLabelled{(a2)}{135}{$7$};
\singleLegLabelled{(a3)}{45}{$1$};
\end{tikzpicture}
\hspace{5pt}-\hspace{5pt}\begin{tikzpicture}[scale=\figScale,baseline=-1.00]
\coordinate (v0) at (0,0);
\foreach\a in {0,...,4}{\coordinate (a\a) at ($(v0)+(\a*\rotn-90-\rotn/2:\edgeLen/2)$);};
\foreach\a[remember=\a as \la] in {0,...,4}{\ifthenelse{\a=0}{}{\draw[int](a\la)--(a\a)coordinate (e\a) at ($(a\la)!0.5!(a\a)$);}};
\singleLegLabelled{(a0)}{-45}{$6$};
\generalCornerMassiveLabelled{(a1)}{225}{$7$}{$3$};
\singleLegLabelled{(a2)}{135}{$4$};
\singleLegLabelled{(a3)}{45}{$5$};
\end{tikzpicture}
\hspace{5pt}+\hspace{5pt}
\begin{tikzpicture}[scale=\figScale,baseline=-1.00]
\coordinate (v0) at (0,0);
\foreach\a in {0,...,4}{\coordinate (a\a) at ($(v0)+(\a*\rotn-90-\rotn/2:\edgeLen/2)$);};
\foreach\a[remember=\a as \la] in {0,...,4}{\ifthenelse{\a=0}{}{\draw[int](a\la)--(a\a)coordinate (e\a) at ($(a\la)!0.5!(a\a)$);}};
\singleLegLabelled{(a0)}{-45}{$2$};
\singleLegLabelled{(a1)}{225}{$3$};
\generalCornerMassiveLabelled{(a2)}{135}{$4$}{$5$};
\generalCornerMassiveLabelled{(a3)}{45}{$6$}{$1$};
\end{tikzpicture}
\hspace{5pt}-\hspace{5pt}\begin{tikzpicture}[scale=\figScale,baseline=-1.00]
\coordinate (v0) at (0,0);
\foreach\a in {0,...,4}{\coordinate (a\a) at ($(v0)+(\a*\rotn-90-\rotn/2:\edgeLen/2)$);};
\foreach\a[remember=\a as \la] in {0,...,4}{\ifthenelse{\a=0}{}{\draw[int](a\la)--(a\a)coordinate (e\a) at ($(a\la)!0.5!(a\a)$);}};
\singleLegLabelled{(a0)}{-45}{$3$};
\singleLegLabelled{(a1)}{225}{$4$};
\generalCornerMassiveLabelled{(a2)}{135}{$5$}{$6$};
\generalCornerMassiveLabelled{(a3)}{45}{$7$}{$2$};
\end{tikzpicture}\,.}
With the conformal regulator, this combination of divergent boxes \emph{happens} to be finite:
\eq{I_{\text{spurious}}^\delta=\frac{1}{2}\log\!\left(\!\frac{\x{1}{3}\x{4}{6}}{\x{1}{4}\x{3}{6}}\!\right)\log\!\left(\!\frac{\x{3}{5}\x{2}{6}\x{1}{4}}{\x{3}{6}\x{2}{4}\x{1}{5}}\!\right)+\mathcal{O}(\delta)\,;}
The fact that this combination is not \emph{truly} finite is clear from its dimensionally-regulated expression:
\eq{I_{\text{spurious}}^{\text{dim-reg}}=\frac{1}{\epsilon}\left[\log\!\left(\!\frac{\x{1}{4}\x{2}{6}\x{3}{7}\x{4}{6}\x{5}{7}}{\x{1}{3}\x{1}{5}\x{2}{7}\x{3}{6}\x{4}{7}}\!\right)+\frac{1}{2}\log\!\left(\!\frac{\x{1}{5}\x{3}{5}\x{4}{6}}{\x{1}{3}\x{2}{4}\x{2}{6}}\!\right)\right]+\mathcal{O}(\epsilon^0)\,.}
This reflects the fact that this combination of boxes does not satisfy the local finiteness criterion given above, as can be seen by noting that e.g.\ only the first box in (\ref{spurious_boxes}) has support in the collinear region associated with leg $1$. The observation that a particular combination of integrals is finite in \emph{some} regularization scheme is not particularly meaningful.\\

The extension to two loops is relatively straightforward: infrared divergences are associated with regions of loop momentum space where one or both loop momenta are collinear to massless external momenta, with the most divergent regions corresponding to additional soft-collinear singularities. A \emph{locally finite} two-loop integrand must vanish in every collinear region of the form
\eq{\label{eq:coll_description_2loop} \ell_1=\alpha\, x_a+(1-\alpha)\,x_{a+1},\quad \ell_2=\beta\, x_b+(1-\beta)\,x_{b+1}.}
We will use the shorthand that a loop momentum in the collinear region associated with leg $a$ satisfies $\ell_i\in\text{coll}(a)$. The \emph{maximal} degrees of divergence associated to particular collinear and soft-collinear residues follow almost trivially from the above one-loop discussion. When the two loops are sent to different collinear regions, i.e.\ $a\neq b$, non-zero residues associated to double-soft-collinear singularities lead to $\log^4(\delta)$ or `one-loop square' divergences in the integrated result. An example of an integrand (which is not relevant for the representation of the ratio function we consider below) with a $\log^4(\delta)$ divergence associated to the double-soft-collinear singularity where $\ell_1=x_1$, $\ell_2=x_4$ is the scalar double-box
\eq{\label{two_loop_double_box_eg}\def\rotn{-90}\def\edgeLen{1*\figScale}\begin{tikzpicture}[scale=\figScale,baseline=0.0]
\coordinate (a0) at (0,0);
\coordinate (a1) at ($(a0)+(180:\edgeLen*0.75)$);
\coordinate (a2) at ($(a1)+(90:\edgeLen*0.75)$);
\coordinate (a3) at ($(a2)+(0:\edgeLen*0.75)$);
\coordinate (a4) at ($(a3)+(0:\edgeLen*0.75)$);
\coordinate (a5) at ($(a0)+(0:\edgeLen*0.75)$);
\draw[int] (a0)--(a1)--(a2)--(a3)--(a4)--(a5)--(a0)--(a3);
\singleLegLabelled{(a0)}{-90}{$5$};
\singleLegLabelled{(a1)}{225}{$6$};
\singleLegLabelled{(a2)}{135}{$1$};
\singleLegLabelled{(a3)}{90}{$2$};
\singleLegLabelled{(a4)}{45}{$3$};
\singleLegLabelled{(a5)}{-45}{$4$};
\end{tikzpicture}\,.}
Indeed, it may be readily verified that this integral is $\log^4(\delta)$-divergent in the conformal-regularization scheme \cite{Bourjaily:2019jrk}. If one of the corners of this topology were to be massive as in
\vspace{-5pt}\eq{\label{two_loop_double_box_eg2}\def\rotn{-90}\def\edgeLen{1*\figScale}\begin{tikzpicture}[scale=\figScale,baseline=0.0]
\coordinate (a0) at (0,0);
\coordinate (a1) at ($(a0)+(180:\edgeLen*0.75)$);
\coordinate (a2) at ($(a1)+(90:\edgeLen*0.75)$);
\coordinate (a3) at ($(a2)+(0:\edgeLen*0.75)$);
\coordinate (a4) at ($(a3)+(0:\edgeLen*0.75)$);
\coordinate (a5) at ($(a0)+(0:\edgeLen*0.75)$);
\draw[int] (a0)--(a1)--(a2)--(a3)--(a4)--(a5)--(a0)--(a3);
\singleLegLabelled{(a0)}{-90}{$5$};
\singleLegLabelled{(a5)}{-45}{$4$};
\singleLegLabelled{(a2)}{135}{$1$};
\singleLegLabelled{(a3)}{90}{$2$};
\singleLegLabelled{(a4)}{45}{$3$};
\generalCornerMassiveLabelled{(a1)}{225}{$6$}{$7$};
\end{tikzpicture}\,,}
then the most divergent configuration would send just one loop soft-collinear, $\ell_2=x_3$, with $\ell_1$ taken in the collinear region associated with leg $1$. This integral would therefore be $\log^3(\delta)$-divergent; adding another mass to one of the massless corners on the right loop would reduce the maximal degree of divergence to $\logsquarediv$, and so-on. 
 
Sub-leading $\logsquarediv$ and $\logdiv$ divergences are also associated with a special `overlapping' collinear region where in (\ref{eq:coll_description_2loop}) we set $a{=}b$---where both loops are collinear with the same massless external leg. Clearly, such divergent regions can be accessible for non-planar integrals only. In fact, in the `almost finite' representation of the two-loop ratio function of \cite{Bourjaily:2015jna}, which we review in more detail in section~\ref{subsec:old_rep_of_ratio_function}, it turns out that such overlapping collinear regions directly correspond to the only remaining local divergences. 

The overlapping region is qualitatively distinct from the generic case because, on the support of the composite cut $\x{\ell_1}{a}=\x{\ell_1}{a{+}1}=\x{\ell_2}{a}=\x{\ell_2}{a{+}1}=0$, the internal propagator vanishes as well, as $\x{\ell_1}{\ell_2}\sim\x{a}{a{+}1}=0$. Because of this, double- and single-soft-collinear  singularities are associated with $\logsquarediv$ and $\logdiv$ divergences, respectively.

\vspace{-0pt}\subsection{Classification of Local Divergences of `Merger' Integrands}\label{subsec:classification_of_ints}\vspace{-0pt}
The representation of the two-loop ratio function given in \cite{Bourjaily:2015jna}---reviewed below near (\ref{ratio_function_explosion})---is written in terms of `mergers' of specific one-loop integrands involving the point at infinity, denoted `$X$', in dual momentum space. The merge operation was defined in \cite{Bourjaily:2015jna} for two $X$-dependent chiral boxes as 
\eq{\begin{split}\mathcal{I}_L\x{\ell_1}{X}\merge\mathcal{I}_R\x{\ell_2}{X}&\equivR \left[\mathcal{I}'_L(\ell_1)\frac{\x{N_L}{X}}{\x{\ell_1}{X}}\right]\merge\left[\mathcal{I}'_R(\ell_2)\frac{\x{N_R}{X}}{\x{\ell_2}{X}}\right]\\ &\equivR\mathcal{I}'_L(\ell_1)\frac{\x{N_L}{N_R}}{\x{\ell_1}{\ell_2}}\mathcal{I}'_R(\ell_2)\,,\end{split}}
where (implicit) symmetrization with respect to loop-momentum labels ensures the merge operation is itself symmetric. To organize the cancellation of infrared singularities in the two-loop ratio function, it suffices to identify the location of all possible divergences in the merger of two (arbitrary) chiral boxes, $\mathcal{I}^i_{a,b,c,d}(\ell_1,X)\otimes\mathcal{I}^j_{e,f,g,h}(\ell_2,X)$. By construction, every chiral box integrand is `one-loop finite' in the sense of section~\ref{subsec:exempli_gratia}. Thus, \emph{every} merger of two such integrands will trivially vanish when the two loops go to \emph{distinct} collinear regions, e.g.\ $\ell_1\in\text{coll}(a)$ and $\ell_2\in\text{coll}(b)$, for $a\neq b$. The remaining divergent regions are those when both loops go to the \emph{same} collinear region, $\ell_1,\ell_2\in\text{coll}(a)$. Clearly, in order for an integrand to have support on this co-dimension-six residue, it must have (at least) one massless corner where leg $a$ sits, for both loops. We shall find a graphical representation of the merger integrands, where the chirality of each box can be labeled by coloring its three-point vertices either white (for $\overline{\text{MHV}}$) or blue (for MHV), to be particularly useful for keeping track of potential sources of divergence. Recalling that the merger of two boxes of the same chirality vanishes, we may write the potentially divergent mergers as all integrands of the form,
\eq{
\label{generic_div_merger}\drawMerger{\oneLoopChiralBoxInt{3}{3}{3}{1}{}{}{}{$a$}{1}}{\oneLoopChiralBoxInt{3}{3}{3}{1}{}{}{}{$a$}{2}}
}
where, throughout this work, we use a `dashed' wedge to indicate an arbitrary leg range, that is, either massless or massive (but not empty), to include all necessary degenerations. Strictly massive legs, on the other hand, will be drawn with a solid wedge, while strictly massless legs will  be drawn with a single leg, as has been done already in e.g. (\ref{spurious_boxes}) and (\ref{two_loop_double_box_eg2}) above. Let us consider each possible leg distribution in (\ref{generic_div_merger}) and locate all sources of infrared divergences. 

While the generic merger of two three-mass boxes which share a massless leg $a$ does not arise in the ratio function, as a simple application of the above discussion we note that it is indeed $\logdiv$-divergent, with non-vanishing support on the co-dimension six residue $\ell_1,\ell_2\in\text{coll}(a)$, i.e. 
\eq{\label{generic_div_merger_with_cuts}\fig{-25pt}{1}{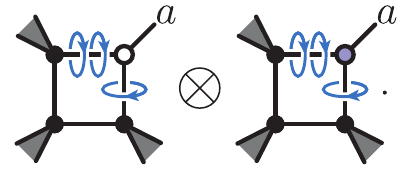}}
Of course, this integrand has support on the additional cuts where we localize $\ell_1,\ell_2$ completely by cutting additional propagators. However, these do not correspond to soft singularities, and as such do not lead to enhanced divergences in the integrated result. In general, we refer to such co-dimension eight residues accessible from the doubly-collinear region as \emph{collinear-but-not-soft}. In fact, mergers involving three and two-mass-easy boxes---in which there are no consecutive massless legs---are also free of any soft singularities, and fit into the same classification according to infrared structure. As such, we may write this class of mergers succinctly as
\eq{\label{coll_but_not_soft_ints}\text{collinear-but-not-soft $\logdiv$:}\quad\drawMerger{\oneLoopChiralBoxInt{2}{3}{2}{1}{}{}{}{$a$}{1}}{\oneLoopChiralBoxInt{2}{3}{2}{1}{}{}{}{$a$}{2}}\,.}

The next class of mergers involves a three- and two-mass-hard box (as well as degenerations). Any merger with (at least) one box with two consecutive massless legs allows access to the soft-collinear singularity in the corresponding loop variable. Single-soft-collinear singularities generate $\logdiv$ divergences upon integration, and all relevant cases can be grouped as the following (including degenerations): 
\eq{\label{all_single_soft_ints}\text{soft-collinear $\logdiv$:}\hspace{10pt}\drawMerger{\oneLoopChiralBoxInt{2}{3}{2}{1}{}{}{}{$a$}{1}}{\oneLoopChiralBoxInt{1}{3}{2}{1}{}{}{}{$a$}{3}}\quad ,\quad \drawMerger{\oneLoopChiralBoxInt{1}{3}{2}{1}{}{}{}{$a$}{4}}{\oneLoopChiralBoxInt{2}{3}{2}{1}{}{}{}{$a$}{2}}.}

Finally, there are those mergers with consecutive massless legs in \emph{both} boxes. These mergers have support on double-soft-collinear singularities, and lead to $\logsquarediv$ divergences upon integration:
\eq{\label{all_double_soft_ints}\text{double-soft-collinear $\logsquarediv$:}\quad\drawMerger{\oneLoopChiralBoxInt{1}{3}{2}{1}{}{}{}{$a$}{4}}{\oneLoopChiralBoxInt{1}{3}{2}{1}{}{}{}{$a$}{3}}\,.}
In section~\ref{subsec:locally_divergent_regions} we will make use of this classification to organize contributions to and confirm the cancellation of all local divergences in the two-loop ratio function of planar sYM.

\newpage
\vspace{-0pt}\section{Local-Finiteness of the Two-Loop Ratio Function in sYM}\label{sec:main_result}\vspace{-0pt}
%

\vspace{-0pt}\subsection{A Nearly-Finite Representation of the Two-Loop Ratio Function}\label{subsec:old_rep_of_ratio_function}\vspace{-0pt}
%
To examine the local finiteness of the two-loop ratio function in planar sYM, the local integrand-level representations for all one and two-loop amplitudes introduced in \cite{Bourjaily:2015jna}---and implemented in the attached \textsc{Mathematica} package `\texttt{two\_loop\_amplitudes}' (see also \cite{Bourjaily:2010wh,Bourjaily:2012gy,Bourjaily:2013mma})---are a particularly good starting point as they make the infrared structure of the amplitude building blocks entirely manifest. By tailoring a basis of integrands to match a minimal set of independent on-shell data sufficient to fix the full amplitude integrand, the one- and two-loop amplitudes admit convenient separations into \emph{manifestly} finite and divergent parts. At one loop, the chiral box expansion reads \cite{Bourjaily:2013mma}:
\eq{\mathcal{A}_{n}^{(k),1}=\overbrace{\sum_{a,b,c,d}\Big({\color{cut1}f_{a,b,c,d}^1}{\color{cut1}\mathcal{I}_{a,b,c,d}^1}\,\pl{\color{cut2}f_{a,b,c,d}^2}{\color{cut2}\mathcal{I}_{a,b,c,d}^2}\Big)\!}^{\text{{\footnotesize$\displaystyle\fwboxR{0pt}{\equivL\,}\mathcal{A}_{n,\text{fin}}^{(k),1}$}}}\pl\overbrace{\mathcal{A}_{n}^{(k),0}\mathcal{I}_{\text{div}}\phantom{\Big)}\hspace{-7pt}}^{\text{{\footnotesize$\displaystyle\fwboxR{0pt}{\equivL}\mathcal{A}_{n,\text{div}}^{(k),1}$}}}\,,\label{chiral_box_expansion}}
where: $\mathcal{A}_n^{(k),L}$ denotes the $L$-loop $n$-point N${}^k$MHV amplitude; $f^i_{a,b,c,d}$ denote the familiar `quadruple-cut' on-shell functions associated with the two co-dimension four residues putting four propagators on-shell,
\eq{\x{\ell}{a}=\x{\ell}{b}=\x{\ell}{c}=\x{\ell}{d}=0\quad\bigger{\Leftrightarrow}\quad f_{a,b,c,d}^{i}\equivR\def\rotn{-90}\def\edgeLen{1*\figScale}\begin{tikzpicture}[scale=\figScale,baseline=-1.5] 
\foreach\a in {0,...,4}{\coordinate (a\a) at ($(v0)+(\a*\rotn-90-\rotn/2:\edgeLen/2)$);};
\foreach\a[remember=\a as \la] in {0,...,4}{\ifthenelse{\a=0}{}{\draw[int](a\la)--(a\a)coordinate (e\a) at ($(a\la)!0.5!(a\a)$);}};
\generalCorner{(a0)}{-45};
\generalCorner{(a1)}{225};
\generalCorner{(a2)}{135};
\generalCorner{(a3)}{45};
\node[genAmp] at (a0) {};
\node[genAmp] at (a1) {};
\node[genAmp] at (a2) {};
\node[genAmp] at (a3) {};
\node at (v0) {$i$}; 
\node[xshift=-1*\figScale,yshift=-0.5*\figScale,label={[label distance=4*\labelDist]177.5:{$a$}}] at (a2) {};
\node[xshift=9.65*\figScale,yshift=1.05*\figScale,label={[label distance=2*\labelDist]225:{$d$}}] at (a1) {};
\node[xshift=1.5*\figScale,yshift=3*\figScale,label={[label distance=3*\labelDist]-35:{$c$}}] at (a0) {};
\node[xshift=-5*\figScale,yshift=0.8*\figScale,label={[label distance=3*\labelDist]75:{$b$}}] at (a3) {};
 \end{tikzpicture}\,;}
and the integrands $\mathcal{I}^i_{a,b,c,d}$ are chiral boxes, given schematically as
\eq{\mathcal{I}_{a,b,c,d}^{i}\equivR\def\rotn{-90}\def\edgeLen{1*\figScale}\begin{tikzpicture}[scale=\figScale,baseline=-1.5]
\foreach\a in {0,...,4}{\coordinate (a\a) at ($(v0)+(\a*\rotn-90-\rotn/2:\edgeLen/2)$);};
\foreach\a[remember=\a as \la] in {0,...,4}{\ifthenelse{\a=0}{}{\draw[int](a\la)--(a\a)coordinate (e\a) at ($(a\la)!0.5!(a\a)$);}};
\generalCorner{(a0)}{-45};
\generalCorner{(a1)}{225};
\generalCorner{(a2)}{135};
\generalCorner{(a3)}{45};
\node[ddot] at (a0) {};
\node[ddot] at (a1) {};
\node[ddot] at (a2) {};
\node[ddot] at (a3) {};
\node at (v0) {$i$}; 
\node[xshift=-1*\figScale,yshift=-0.5*\figScale,label={[label distance=4*\labelDist]177.5:{$a$}}] at (a2) {};
\node[xshift=9.65*\figScale,yshift=1.05*\figScale,label={[label distance=2*\labelDist]225:{$d$}}] at (a1) {};
\node[xshift=1.5*\figScale,yshift=3*\figScale,label={[label distance=3*\labelDist]-35:{$c$}}] at (a0) {};
\node[xshift=-5*\figScale,yshift=0.8*\figScale,label={[label distance=3*\labelDist]75:{$b$}}] at (a3) {};
 \end{tikzpicture}\bigger{\Leftrightarrow}\quad\frac{\x{\ell}{N^i}\x{Y^i}{X}}{\x{\ell}{a}\x{\ell}{b}\x{\ell}{c}\x{\ell}{d}\x{\ell}{X}}\,,}
which are designed to have support on one of the associated quadruple cuts and vanish on the other, and where $X\equivR x_\infty$ is the point at infinity in (dual) loop momentum space. Explicit expressions for these integrands given in terms of momentum twistors may be found in \cite{Bourjaily:2015jna}, though their explicit form will not be needed here. Finally, the divergent part of one loop amplitudes is obtained by matching term-by-term all composite leading singularities with `chiral triangle' integrands,
\eq{\mathcal{A}_n^{(k),0}=\hspace{-15pt}\fig{-30pt}{1}{soft_collinear_LS}\bigger{\Leftrightarrow}\;\mathcal{I}_{a}(X)\equivR\frac{\x{X}{a}\x{a{-}1}{a{+}1}}{\x{\ell}{a{-}1}\x{\ell}{a}\x{\ell}{a{+}1}\x{\ell}{X}}\,,\;\text{and}\;\;\;\mathcal{I}_{\text{div}}\equivR\!\sum_a\mathcal{I}_a(X)\,.}
In the expansion (\ref{chiral_box_expansion}), each chiral box integrand is dressed with the appropriate $\text{N}^k$MHV on-shell function. Because the colors and labels of an integrand and its associated on-shell function coefficient are identical, we shall find convenient a diagrammatic notation which combines both the on-shell function and the chiral box integrand multiplying it in (\ref{chiral_box_expansion}) into a single figure using \emph{spherical} (colored---to indicate N${}^k$MHV degree) vertices:
\eq{\def\rotn{-90}\def\edgeLen{1*\figScale}\begin{tikzpicture}[scale=\figScale,baseline=-1.5]
\foreach\a in {0,...,4}{\coordinate (a\a) at ($(v0)+(\a*\rotn-90-\rotn/2:\edgeLen/2)$);};
\foreach\a[remember=\a as \la] in {0,...,4}{\ifthenelse{\a=0}{}{\draw[int](a\la)--(a\a)coordinate (e\a) at ($(a\la)!0.5!(a\a)$);}};
\generalCorner{(a0)}{-45};
\generalCorner{(a1)}{225};
\generalCorner{(a2)}{135};
\generalCorner{(a3)}{45};
\node[intTermCornerGen] at (a0) {};
\node[intTermCornerGen] at (a1) {};
\node[intTermCornerGen] at (a2) {};
\node[intTermCornerGen] at (a3) {};
\node at (v0) {$i$}; 
\node[xshift=-1*\figScale,yshift=-0.5*\figScale,label={[label distance=4*\labelDist]177.5:{$a$}}] at (a2) {};
\node[xshift=9.65*\figScale,yshift=1.05*\figScale,label={[label distance=2*\labelDist]225:{$d$}}] at (a1) {};
\node[xshift=1.5*\figScale,yshift=3*\figScale,label={[label distance=3*\labelDist]-35:{$c$}}] at (a0) {};
\node[xshift=-5*\figScale,yshift=0.8*\figScale,label={[label distance=3*\labelDist]75:{$b$}}] at (a3) {};
\end{tikzpicture} \equivR \begin{tikzpicture}[scale=\figScale,baseline=-1.5] 
\foreach\a in {0,...,4}{\coordinate (a\a) at ($(v0)+(\a*\rotn-90-\rotn/2:\edgeLen/2)$);};
\foreach\a[remember=\a as \la] in {0,...,4}{\ifthenelse{\a=0}{}{\draw[int](a\la)--(a\a)coordinate (e\a) at ($(a\la)!0.5!(a\a)$);}};
\generalCorner{(a0)}{-45};
\generalCorner{(a1)}{225};
\generalCorner{(a2)}{135};
\generalCorner{(a3)}{45};
\node[genAmp] at (a0) {};
\node[genAmp] at (a1) {};
\node[genAmp] at (a2) {};
\node[genAmp] at (a3) {};
\node at (v0) {$i$}; 
\node[xshift=-1*\figScale,yshift=-0.5*\figScale,label={[label distance=4*\labelDist]177.5:{$a$}}] at (a2) {};
\node[xshift=9.65*\figScale,yshift=1.05*\figScale,label={[label distance=2*\labelDist]225:{$d$}}] at (a1) {};
\node[xshift=1.5*\figScale,yshift=3*\figScale,label={[label distance=3*\labelDist]-35:{$c$}}] at (a0) {};
\node[xshift=-5*\figScale,yshift=0.8*\figScale,label={[label distance=3*\labelDist]75:{$b$}}] at (a3) {};
 \end{tikzpicture} \times \begin{tikzpicture}[scale=\figScale,baseline=-1.5]
\foreach\a in {0,...,4}{\coordinate (a\a) at ($(v0)+(\a*\rotn-90-\rotn/2:\edgeLen/2)$);};
\foreach\a[remember=\a as \la] in {0,...,4}{\ifthenelse{\a=0}{}{\draw[int](a\la)--(a\a)coordinate (e\a) at ($(a\la)!0.5!(a\a)$);}};
\generalCorner{(a0)}{-45};
\generalCorner{(a1)}{225};
\generalCorner{(a2)}{135};
\generalCorner{(a3)}{45};
\node[ddot] at (a0) {};
\node[ddot] at (a1) {};
\node[ddot] at (a2) {};
\node[ddot] at (a3) {};
\node at (v0) {$i$}; 
\node[xshift=-1*\figScale,yshift=-0.5*\figScale,label={[label distance=4*\labelDist]177.5:{$a$}}] at (a2) {};
\node[xshift=9.65*\figScale,yshift=1.05*\figScale,label={[label distance=2*\labelDist]225:{$d$}}] at (a1) {};
\node[xshift=1.5*\figScale,yshift=3*\figScale,label={[label distance=3*\labelDist]-35:{$c$}}] at (a0) {};
\node[xshift=-5*\figScale,yshift=0.8*\figScale,label={[label distance=3*\labelDist]75:{$b$}}] at (a3) {};
 \end{tikzpicture}\,.}
As usual, we shall use blue (white) vertices to indicate both the respective MHV ($\overline{\text{MHV}}$) amplitudes in the on-shell function \emph{and} the corresponding quad-cut solution on which the integrand has support. As an example, this notation allows us to write the MHV one-loop amplitude integrand as a sum of two-mass-easy boxes (with one-mass degenerations included in the sum):
 \eq{\mathcal{A}^{(0),1}_n=\sum_{a<b}\hspace{5pt}\oneLoopMHVInt{3}{1}{2}{1}{}{$b$}{}{$a$}+\mathcal{A}_n^{(0),0}\mathcal{I}_{\text{div}}\,.\label{mhv_one_loop}}
(Of course, for MHV amplitudes this notation is entirely unnecessary, for the simple reason that every on-shell function in this case is equal to the tree-level amplitude, i.e. in momentum twistor space $\mathcal{A}^{(0),0}_n=1$.) 

Because the divergent piece of (\ref{chiral_box_expansion}) is universal, this representation trivially yields a manifestly, locally-finite representation of the one-loop ratio function,
\eq{\begin{split}\mathcal{R}_{n}^{(k),1}&= {\color{hred}\Big[}\mathcal{A}_{n}^{(k),1}{\color{hred}\Big]}\mi\mathcal{A}_{n}^{(k),0}{\color{hblue}\Big[}\mathcal{A}_{n}^{(0),1}{\color{hblue}\Big]}\,,\\
&={\color{hred}\Big[}\mathcal{A}_{n,\text{fin}}^{(k),1}\pl \mathcal{A}_{n}^{(k),0}\mathcal{I}_{\text{div}}{\color{hred}\Big]}\mi\mathcal{A}_{n}^{(k),0}{\color{hblue}\Big[}\mathcal{A}_{n,\text{fin}}^{(0),1}\pl \mathcal{A}_{n}^{(0),0}\mathcal{I}_{\text{div}}{\color{hblue}\Big]}\,,\\
&=\mathcal{A}_{n,\text{fin}}^{(k),1}\mi\mathcal{A}_{n}^{(k),0}\mathcal{A}_{n,\text{fin}}^{(0),1}\,.
\end{split}\label{one_loop_ratio_finiteness}}

At two loops, there is a similar separation of finite and divergent pieces, 
\eq{\mathcal{A}_{n}^{(k),2}\equivR\mathcal{A}_{n,\text{fin}}^{(k),2}\pl\mathcal{A}_{n,\text{div}}^{(k),2}\,,\quad\text{with}\quad\mathcal{A}_{n,\text{div}}^{(k),2}\equivR\mathcal{A}_{n,\text{fin}}^{(k),1}\merge\mathcal{I}_{\text{div}}\pl\mathcal{A}_{n}^{(k),0}\frac{1}{2}\Big(\mathcal{I}_{\text{div}}\merge\mathcal{I}_{\text{div}}\Big)\label{general_two_loop_amplitude_representation}}
where we have written the divergent part of the amplitude in terms of the `merge' operation introduced above. (For a more thorough discussion of the finite part of the amplitude, see \cite{Bourjaily:2015jna}.) The two-loop ratio function is then defined as
\eq{\begin{split}
&\mathcal{R}_n^{(k),2}\equivR\mathcal{A}_{n}^{(k),2}\mi\mathcal{A}_{n}^{(k),1}\!\!\times\!\mathcal{A}_{n}^{(0),1}\mi\mathcal{A}_{n}^{(k),0}\!\!\times\!\!\Big(\mathcal{A}_{n}^{(0),2}\mi\mathcal{A}_{n}^{(0),1}\!\!\times\!\mathcal{A}_{n}^{(0),1}\Big)\,.\end{split}\label{two_loop_ratio_definition}}
Substituting the integrand-level expressions (\ref{general_two_loop_amplitude_representation}) and (\ref{chiral_box_expansion}), using the fact that in momentum twistor space, $\mathcal{A}_n^{(0),0}=1$, and the convenient fact proven in \cite{Bourjaily:2015jna} that (for entire amplitude integrands) $\mathcal{A}_n^{(k_1),1}\!\!\times\mathcal{A}_n^{(k_2),1}=\mathcal{A}_n^{(k_1),1}\merge\mathcal{A}_n^{(k_2),1}$, we immediately obtain an `almost finite' representation of the two-loop ratio function:
\vspace{-4pt}\eq{\begin{split}\hspace{-50pt}\mathcal{R}_n^{(k),2}\hspace{-0pt}&=\phantom{\mi}\phantom{{\color{hred}\Big[}}\mathcal{A}_{n}^{(0),2}\phantom{{\color{hred}\Big]}}\mi\phantom{{\color{hblue}\Big[}}\mathcal{A}_{n}^{(k),1}\times\mathcal{A}_{n}^{(0),1}\phantom{{\color{hblue}\Big]}}\mi\mathcal{A}_{n}^{(k),0}\phantom{{\color{hteal}\Big[}}\mathcal{A}_{n}^{(0),2}\mi\mathcal{A}_{n}^{(0),1}\times\mathcal{A}_{n}^{(0),1}\phantom{{\color{hteal}\Big]}}\\
&=\phantom{\mi}{\color{hred}\Big[}\mathcal{A}_{n}^{(0),2}{\color{hred}\Big]}\mi{\color{hblue}\Big[}\mathcal{A}_{n}^{(k),1}\merge\mathcal{A}_{n}^{(0),1}{\color{hblue}\Big]}\mi\mathcal{A}_{n}^{(k),0}{\color{hteal}\Big[}\mathcal{A}_{n}^{(0),2}\mi\mathcal{A}_{n}^{(0),1}\merge\mathcal{A}_{n}^{(0),1}{\color{hteal}\Big]}\\
&=\phantom{\mi}{\color{hred}\Big[}\mathcal{A}_{n,\text{fin}}^{(k),2}\hspace{0pt}\pl\mathcal{A}_{n,\text{fin}}^{(k),1}\merge\mathcal{I}_{\text{div}}\hspace{0pt}\pl\frac{1}{2}\mathcal{A}_{n}^{(k),0}\mathcal{I}_{\text{div}}\merge\mathcal{I}_{\text{div}}{\color{hred}\Big]}\\
&\phantom{=}\hspace{0.0pt}\,\,\,\mi{\color{hblue}\Big[}\Big(\mathcal{A}_{n,\text{fin}}^{(k),1}\pl\mathcal{A}_{n}^{(k),0}\mathcal{I}_{\text{div}}\Big)\merge\Big(\mathcal{A}_{n,\text{fin}}^{(0),1}\pl\mathcal{I}_{\text{div}}\Big){\color{hblue}\Big]}\\
&\phantom{=}\hspace{0.0pt}\,\,\,\mi\mathcal{A}_{n}^{(k),0}{\color{hteal}\Big[}\mathcal{A}_{n,\text{fin}}^{(0),2}\hspace{0pt}\pl\mathcal{A}_{n,\text{fin}}^{(0),1}\merge\mathcal{I}_{\text{div}}\hspace{0pt}\pl\frac{1}{2}\mathcal{I}_{\text{div}}\merge\mathcal{I}_{\text{div}}\mi\Big(\mathcal{A}_{n,\text{fin}}^{(0),1}\pl\mathcal{I}_{\text{div}}\Big)\merge\Big(\mathcal{A}_{n,\text{fin}}^{(0),1}\pl\mathcal{I}_{\text{div}}\Big){\color{hteal}\Big]}\hspace{-50pt}\\
&=\phantom{\mi}\mathcal{A}_{n,\mathrm{fin}}^{(k),2}\mi\mathcal{A}_{n,\mathrm{fin}}^{(k),1}\merge\mathcal{A}_{n,\mathrm{fin}}^{(0),1}\mi\mathcal{A}_{n}^{(k),0}\Big(\mathcal{A}_{n,\mathrm{fin}}^{(0),2}\mi\mathcal{A}_{n,\mathrm{fin}}^{(0),1}\merge\mathcal{A}_{n,\mathrm{fin}}^{(0),1}\Big)\,.\\[-22pt]~\vspace{-0pt}\end{split}\label{ratio_function_explosion}\vspace{-10pt}}
While the merger of two `one-loop finite' integrands does not, in general, yield an integrand which is locally finite at two loops, the locations of all remaining divergences are essentially trivial to identify: such an integrand can only have support on the overlapping regions where both loop momenta are collinear with the same massless particle. To demonstrate the local finiteness of the ratio function, we must show that all such overlapping collinear divergences present in the two terms $\Big[\mathcal{A}_{n}^{(k),0}\mathcal{A}_{n,\mathrm{fin}}^{(0),1}\merge\mathcal{A}_{n,\mathrm{fin}}^{(0),1}\Big]\mi\Big[\mathcal{A}_{n,\mathrm{fin}}^{(k),1}\merge\mathcal{A}_{n,\mathrm{fin}}^{(0),1}\Big]$ cancel.

\newpage

\vspace{-0pt}\subsection{The Local Divergence of the MHV$\times$MHV Contribution}\label{subsec:mhv_times_mhv_soft_collinear_div}\vspace{-0pt}
To demonstrate the cancellation of all overlapping collinear divergences appearing term-by-term in the representation of the two-loop ratio function (\ref{ratio_function_explosion}), it is useful to examine the divergent part of the MHV$\times$MHV contribution separately first. The one-loop, finite part of the MHV amplitude integrand $\mathcal{A}_n^{(0),1}$ is a sum of two-mass easy boxes (and one-mass degenerations) as shown in (\ref{mhv_one_loop}). Consider the local divergence associated with, say, leg $a$, and the corresponding collinear region where the loop momenta are parametrized as
\eq{\label{overlapping_collinear_region}\ell_1=\alpha\,x_a+(1-\alpha)\,x_{a+1},\quad \ell_2=\beta\,x_a+(1-\beta)\,x_{a+1}\,.}
The only mergers appearing in $\mathcal{A}_{n,\mathrm{fin}}^{(0),1}\merge\mathcal{A}_{n,\mathrm{fin}}^{(0),1}$ with support in this region involve a massless leg $a$ in both the left and right boxes; that is, both must be one of the following chiral box integrands:
\eq{\hspace{-5pt}\tikzBox{\oneLoopMHVInt{1}{1}{2}{1}{$a$}{}{}{}}\hspace{25pt}\text{or}\hspace{-10pt}\tikzBox{\oneLoopMHVInt{3}{1}{2}{1}{}{}{}{$a$}}\hspace{25pt}.}
As the merger of two boxes of the same chirality always vanishes, this implies that the full MHV$\times$MHV divergence arising from the collinear region associated with leg $a$ can be written as
\eq{\label{mhv_divergence1} \mathcal{A}_{n,\mathrm{fin}}^{(0),1}\merge\mathcal{A}_{n,\mathrm{fin}}^{(0),1}\bigger{\supset} 2\sum_b\drawMerger{\oneLoopMHVInt{1}{1}{2}{1}{$a$}{}{}{}}{\oneLoopMHVInt{3}{1}{2}{1}{}{$b$}{}{$a$}}}
In fact, the local divergence is even simpler than the representation (\ref{mhv_divergence1}) suggests. If we consider the residue associated with setting $\x{\ell_i}{b}\rightarrow0$, where $b$ is generic, there are two contributing terms in (\ref{mhv_divergence1}) which cancel pairwise: 
\eq{\label{mhv_cancellation_eg}\underset{\substack{\ell_1,\ell_2\in\mathrm{coll}(a) \\ \x{\ell_i}{b}=0}}{\Res}\mathcal{A}_{n,\mathrm{fin}}^{(0),1}\merge\mathcal{A}_{n,\mathrm{fin}}^{(0),1}=2\,\fig{-22.5pt}{1}{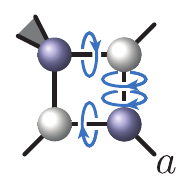}\merge\left(\fig{-22.5pt}{1}{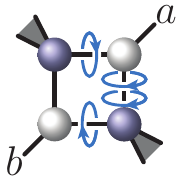}+\fig{-22.5pt}{1}{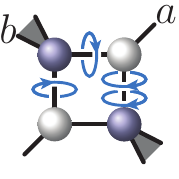}\right)=0\,.}
The same pairwise cancellation works out for every collinear-but-not-soft pole. As a consequence, the residue (\ref{mhv_divergence1}) in the collinear region is \emph{purely} double-soft-collinear, and in terms of the parametrization (\ref{overlapping_collinear_region}) can be written as a two-form in $\alpha,\beta$, 
\eq{\label{mhv_hexacut}\underset{\ell_1,\ell_2\in\mathrm{coll}(a)}{\Res}\mathcal{A}_{n,\mathrm{fin}}^{(0),1}\merge\mathcal{A}_{n,\mathrm{fin}}^{(0),1}=\frac{2\,d\alpha\,d\beta}{\alpha(\alpha-1)\beta(\beta-1)}\,.}
This implies that in the two-loop ratio function, all local divergences that are \emph{not} of the double-soft-collinear type must cancel amongst the $\text{N}^k$MHV$\times$MHV terms themselves.

\newpage
\vspace{-0pt}\subsection{Cancellation of All Local Divergences in the Ratio Function}\label{subsec:locally_divergent_regions}\vspace{-0pt}
%
The remaining divergent terms in the representation of the ratio function given in (\ref{ratio_function_explosion}) are written in terms of the mergers of the finite parts of $\text{N}^k\text{MHV}$ and MHV one-loop integrands. Every potentially divergent merger was classified in subsection~\ref{subsec:classification_of_ints}, so we may organize the divergent contributions which must cancel according to the three classes of divergent integrands introduced there. From the result of the previous subsection, it follows that the double-soft-collinear $\logsquarediv$ divergences in $\text{N}^k$MHV$\times$MHV must combine to give twice the tree amplitude, while all other local divergences---those associated with singly-soft-collinear and collinear-but-not-soft singularities---must cancel separately.

\vspace{-0pt}\subsubsection{Soft-Collinear $\logsquarediv$-Divergent Regions}\label{subsubsec:log2_div_regions}\vspace{-0pt}
%
Let us consider the double-soft-collinear singularity where $\ell_1=\ell_2=x_a$. From the general classification of subsection~\ref{subsec:classification_of_ints}, every non-zero merger with two consecutive massless legs $a{-}1, a$ in both left and right boxes appearing in the ratio function have these divergences. For the MHV$\times$MHV part of the ratio function, there is a single such merger (which contributes with multiplicity two),
\eq{\label{mhv2_log2}\underset{\ell_1,\ell_2{=}x_a}{\Res}\mathcal{A}_{n}^{(k),0}\mathcal{A}_{n,\mathrm{fin}}^{(0),1}\merge\mathcal{A}_{n,\mathrm{fin}}^{(0),1}=2\,\mathcal{A}_{n}^{(k),0}\times\fig{-22.5pt}{1}{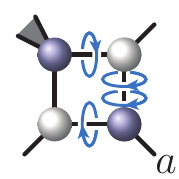}\merge\fig{-22.5pt}{1}{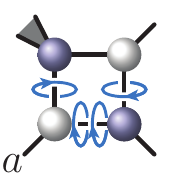}=2\,\mathcal{A}_{n}^{(k),0}.}
Alternatively, this follows from the residue of the MHV$\times$MHV mergers on the full hexa-cut surface (\ref{mhv_hexacut}). For the $\text{N}^k$MHV$\times$MHV terms, there are six merger topologies of the general form~(\ref{all_double_soft_ints}). Adopting the convention that every $\text{N}^k$MHV box is drawn on the left-hand-side of the merger symbol $\merge$, the integrands with this local divergence may be written as 
\eq{\drawMerger{\oneLoopLSxInt{1}{3}{2}{1}{$a$}{}{}{}{4}}{\oneLoopLSxInt{1}{1}{2}{1}{}{}{}{$a$}{10}}\quad,\quad \drawMerger{\oneLoopLSxInt{1}{3}{2}{1}{$a$}{}{}{}{3}}{\oneLoopLSxInt{1}{1}{2}{1}{$a$}{}{}{}{11}}\,.}
Conveniently, we may group the terms in the $\text{N}^k$MHV$\times$MHV part of the ratio function with support on the double-soft-collinear residue $\ell_1=\ell_2=x_a$ into two `BCFW' sets \cite{BCF} according to:
\eq{\begin{split}\hspace{-40pt}\label{log2bcfwSets}\underset{\substack{{\ell_1=x_a} \\ {\ell_2=x_a}}}{\Res}\mathcal{A}_{n,\mathrm{fin}}^{(k),1}\merge\mathcal{A}_{n,\mathrm{fin}}^{(0),1}=\phantom{\!+\!}\Bigg(&\fig{-22.5pt}{1}{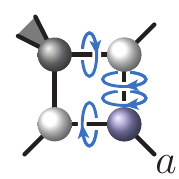}+\fig{-22.5pt}{1}{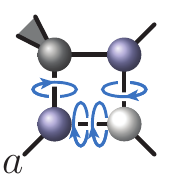}+\sum\fig{-22.5pt}{1}{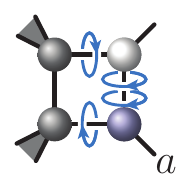}\Bigg)\merge\fig{-22.5pt}{1}{log2_mhv2} \hspace{-36pt}\\\!+\!\Bigg(&\fig{-22.5pt}{1}{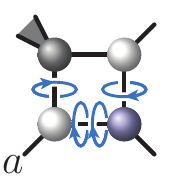}+\fig{-22.5pt}{1}{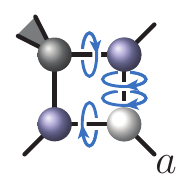}+\sum\fig{-22.5pt}{1}{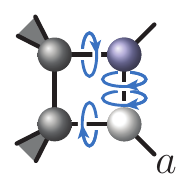}\Bigg)\merge\fig{-22.5pt}{1}{log2_mhv1}\,.\hspace{-36pt}\end{split}}
As each integrand merger in this expression is unit on the double-soft-collinear singularity $\ell_1=\ell_2=x_a$, each factor in parentheses in (\ref{log2bcfwSets}) becomes a particular BCFW recursion of the $\text{N}^k$MHV tree amplitude, combining to give exactly a total contribution of $-2\,\mathcal{A}_{n}^{(k),0}$ to the ratio function; this precisely cancels against the `MHV$\times$MHV' contribution (\ref{mhv2_log2}). 

The above argument was for the double-soft-collinear singularity where $\ell_1=\ell_2=x_a$. The local divergence associated with the double-soft-collinear singularity where the loops approach different dual points, e.g. $\ell_1=x_a$ and $\ell_2=x_{a+1}$, cancels by an analogous grouping of the relevant integrands into BFCW sets. 

\vspace{-0pt}\subsubsection{Soft-Collinear $\logdiv$-Divergent Regions}\label{subsubsec:log1_div_regions}\vspace{-0pt}
%
The cancellation of the $\logdiv$ divergences associated with a single-soft-collinear singularity is, by the general arguments above, among the $\text{N}^k$MHV$\times$MHV terms \emph{themselves}. The set of mergers with single-soft-collinear divergences was given in (\ref{all_single_soft_ints}); however, to organize their cancellation, it is useful to consider the subset of these mergers which have support on a co-dimension eight residue where one loop is soft-collinear, e.g. $\ell_i\rightarrow x_a$, while the other loop is collinear and cuts a non-adjacent propagator, $\ell_j\in\text{coll}(a)$ and $\x{\ell_j}{b}\rightarrow0$. Concretely, we may parametrize the on-shell loop momenta in this case as
\eq{\ell_i=x_a,\quad \ell_j=\beta^\star\, x_a+(1-\beta^\star)\,x_{a+1}\,,}
where $\beta^\star$ is the (unique) solution to $\x{\ell_2}{b}=0$, whose explicit form is not important for our discussion. The cancellation of single-soft-collinear divergences in the ratio function is equivalent to the statement that all such co-dimension eight residues vanish.

There are two kinds of cancellations we must consider, depending on which loop goes soft-collinear. If we consider the divergence associated with sending the $\text{N}^k$MHV loop to the soft-collinear point $x_{a+1}$ and `collecting' on the pole associated with the dual momentum $x_b$, there are exactly two mergers involving the \emph{same} $\text{N}^k$MHV chiral box (and associated on-shell function) which cancel pairwise: 
\eq{\label{single_soft_cancel_type1}\underset{\substack{{\ell_1=x_a} \\ {\ell_2\in\text{coll}(a)} \\ {\x{\ell_2}{b}=0}}}{\Res}\mathcal{A}_{n,\mathrm{fin}}^{(k),1}\merge\mathcal{A}_{n,\mathrm{fin}}^{(0),1}\bigger{\supset}\fig{-22.5pt}{1}{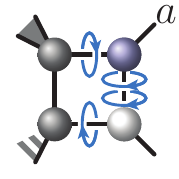}\merge\left(\fig{-22.5pt}{1}{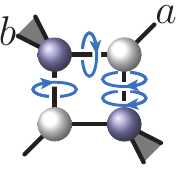}+\fig{-22.5pt}{1}{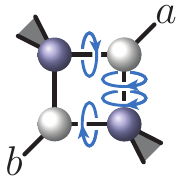}\right)=0\,.}

On the other hand, if we consider going soft-collinear in the MHV loop and collect on poles in the $\text{N}^k$MHV boxes, it is easy to see the cancellation cannot (generically) be pairwise for the simple reason that there are no two-term identities between distinct $\text{N}^k$MHV one-loop on-shell functions. Instead, the cancellation of local divergences arises as a particular instance of Cauchy's residue theorem \cite{GriffithsHarris,Bourjaily:2018omh}. To illustrate this explicitly, let us first consider the terms in the ratio function involving $\text{N}^k$MHV boxes whose on-shell coefficients include a white massless vertex with leg $a$. Schematically, the contributing terms read
\begin{align}\label{single_soft_cancel_type2}\hspace{-10pt}\underset{\substack{{\ell_2=x_a} \\ {\ell_1\in\text{coll}(a)} \\ {\x{\ell_1}{b}=0}}}{\Res}\mathcal{A}_{n,\mathrm{fin}}^{(k),1}\merge\mathcal{A}_{n,\mathrm{fin}}^{(0),1}\bigger{\supset}&\phantom{+}\Bigg(\fig{-22.5pt}{1}{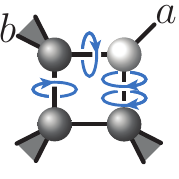}+\fig{-22.5pt}{1}{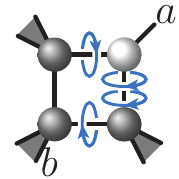}+\fig{-22.5pt}{1}{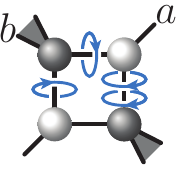} \\ &\hspace{7pt}+\fig{-22.5pt}{1}{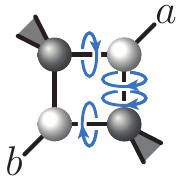}+\fig{-22.5pt}{1}{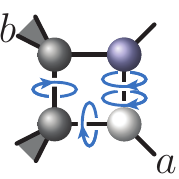}+\fig{-22.5pt}{1}{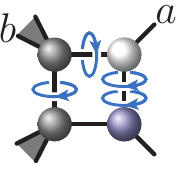}\Bigg)\merge\fig{-22.5pt}{1}{log2_mhv1}\,.\nonumber\end{align}
This expression should be understood as being summed over all the possible leg distributions on the four- or higher-point vertices. It may be immediately verified using e.g. the tools found in \cite{Bourjaily:2013mma} that, for any given choice of helicity degree $k$, the particular sum of on-shell functions appearing in the $\text{N}^k$MHV side of (\ref{single_soft_cancel_type2}) vanishes when the appropriate signs from a residues of the cut integrands are taken into account. In fact, the necessary identity can be directly understood as the residue theorem associated with the boundaries of the chiral two-mass triangle sub-leading singularity in sYM. Namely, letting `$\partial[\cdot]$' denote the set of all maximal co-dimension residues obtained by factorization of a diagram's vertices, then the resulting collection of leading singularities:\\[0pt]
\vspace{-6pt}\eq{\label{grt1}\hspace{-75pt}\bigger{\partial}\!\!\!\left[\rule{0pt}{24pt}\right.\hspace{-8pt}\begin{tikzpicture}[scale=\figScale,baseline=-6.5]\def\rotn{-120}
\coordinate (v0) at (0,0);
\foreach\a in {0,...,3}{\coordinate (a\a) at ($(v0)+(\a*\rotn-30-\rotn/2:\edgeLen/2)$);};
\foreach\a[remember=\a as \la] in {0,...,3}{\ifthenelse{\a=0}{}{\draw[int](a\la)--(a\a)coordinate (e\a) at ($(a\la)!0.5!(a\a)$);}};
\singleLeg{(a0)}{30};
\node[mhvBarAmp] at (a0) {};
\node[label={[label distance=4*\labelDist]+20:{$a$}}] at (a0) {};
\generalCornerMassive{(a1)}{270};
\node[genAmp] at (a1) {};
\generalCornerMassive{(a2)}{150};
\node[genAmp] at (a2) {};
\node[xshift=-3.0*\figScale,yshift=3.0*\figScale,label={[label distance=4*\labelDist]178:{$b$}}] at (a2) {};
\end{tikzpicture}\left.\hspace{-6pt}\rule{0pt}{28pt}\right]\!\!\!=\!\!\left\{\rule{0pt}{28pt}\hspace{4pt}
\begin{tikzpicture}[scale=\figScale,baseline=-1.5]\def\rotn{-90}\def\edgeLen{1*\figScale}
\useasboundingbox ($\figScale*(-0.6,-0.6)$) rectangle ($\figScale*(0.6,0.6)$);
\coordinate (v0) at (0,0);\foreach\a in {0,...,4}{\coordinate (a\a) at ($(v0)+(\a*\rotn-90-\rotn/2:\edgeLen/2)$);};%
\draw[int](a2)--(a3)--(a0)--(a1);\draw[intH](a1)--(a2);\generalCornerMassive{(a0)}{-45};\generalCornerMassive{(a2)}{135};\node[label={[label distance=2*\labelDist]45:{$a$}}] at (a3) {};\node[xshift=3,label={[label distance=2*\labelDist]-135:{$b$}}] at (a1) {};
\generalCornerMassive{(a1)}{-135};\singleLeg{(a3)}{45};
\node[genAmp] at (a2) {};\node[mhvBarAmp] at (a3) {};\node[genAmp] at (a1) {};\node[genAmp] at (a0) {};
\end{tikzpicture}\hspace{7pt},\hspace{6pt}
\begin{tikzpicture}[scale=\figScale,baseline=-1.5]\def\rotn{-90}\def\edgeLen{1*\figScale}
\useasboundingbox ($\figScale*(-0.6,-0.6)$) rectangle ($\figScale*(0.6,0.6)$);
\coordinate (v0) at (0,0);\foreach\a in {0,...,4}{\coordinate (a\a) at ($(v0)+(\a*\rotn-90-\rotn/2:\edgeLen/2)$);};%
\draw[int](a2)--(a3)--(a0)--(a1);\draw[intH](a1)--(a2);\generalCornerMassive{(a0)}{-45};\generalCornerMassive{(a2)}{135};\node[label={[label distance=2*\labelDist]45:{$a$}}] at (a3) {};\node[label={[label distance=2*\labelDist]-135:{$b$}}] at (a1) {};
\singleLeg{(a1)}{-135};\singleLeg{(a3)}{45};
\node[genAmp] at (a2) {};\node[mhvBarAmp] at (a3) {};\node[mhvBarAmp] at (a1) {};\node[genAmp] at (a0) {};
\end{tikzpicture}\hspace{7pt},\hspace{6pt}
\begin{tikzpicture}[scale=\figScale,baseline=-1.5]\def\rotn{-90}\def\edgeLen{1*\figScale}
\useasboundingbox ($\figScale*(-0.6,-0.6)$) rectangle ($\figScale*(0.6,0.6)$);
\coordinate (v0) at (0,0);\foreach\a in {0,...,4}{\coordinate (a\a) at ($(v0)+(\a*\rotn-90-\rotn/2:\edgeLen/2)$);};%
\draw[int](a2)--(a3)--(a0)--(a1);\draw[intH](a1)--(a2);\generalCornerMassive{(a0)}{-45};\generalCornerMassive{(a1)}{-135};\node[label={[label distance=2*\labelDist]45:{$a$}}] at (a3) {};\singleLeg{(a2)}{135};\singleLeg{(a3)}{45};\node[xshift=3,label={[label distance=2*\labelDist]-135:{$b$}}] at (a1) {};
\node[genAmp] at (a1) {};\node[mhvBarAmp] at (a3) {};\node[mhvAmp] at (a2) {};\node[genAmp] at (a0) {};
\end{tikzpicture}\hspace{7pt},\hspace{6pt}
\begin{tikzpicture}[scale=\figScale,baseline=-1.5]\def\rotn{-90}\def\edgeLen{1*\figScale}
\useasboundingbox ($\figScale*(-0.6,-0.6)$) rectangle ($\figScale*(0.6,0.6)$);
\coordinate (v0) at (0,0);\foreach\a in {0,...,4}{\coordinate (a\a) at ($(v0)+(\a*\rotn-90-\rotn/2:\edgeLen/2)$);};%
\draw[int](a1)--(a2)--(a3)--(a0);\draw[intH](a0)--(a1);\generalCornerMassive{(a0)}{-45};\generalCornerMassive{(a2)}{135};\node[label={[label distance=2*\labelDist]45:{$a$}}] at (a3) {};\node[yshift=-7,xshift=-2,label={[label distance=2*\labelDist]135:{$b$}}] at (a2) {};
\generalCornerMassive{(a1)}{-135};\singleLeg{(a3)}{45};
\node[genAmp] at (a2) {};\node[mhvBarAmp] at (a3) {};\node[genAmp] at (a1) {};\node[genAmp] at (a0) {};
\end{tikzpicture}\hspace{7pt},\hspace{6pt}
\begin{tikzpicture}[scale=\figScale,baseline=-1.5]\def\rotn{-90}\def\edgeLen{1*\figScale}
\useasboundingbox ($\figScale*(-0.6,-0.6)$) rectangle ($\figScale*(0.6,0.6)$);
\coordinate (v0) at (0,0);\foreach\a in {0,...,4}{\coordinate (a\a) at ($(v0)+(\a*\rotn-90-\rotn/2:\edgeLen/2)$);};%
\draw[int](a1)--(a2)--(a3)--(a0);\draw[intH](a0)--(a1);\generalCornerMassive{(a0)}{-45};\generalCornerMassive{(a2)}{135};\node[label={[label distance=2*\labelDist]45:{$a$}}] at (a3) {};\node[yshift=-7,xshift=-2,label={[label distance=2*\labelDist]135:{$b$}}] at (a2) {};
\singleLeg{(a1)}{-135};\singleLeg{(a3)}{45};
\node[genAmp] at (a2) {};\node[mhvBarAmp] at (a3) {};\node[mhvBarAmp] at (a1) {};\node[genAmp] at (a0) {};
\end{tikzpicture}\hspace{7pt},\hspace{6pt}
\begin{tikzpicture}[scale=\figScale,baseline=-1.5]\def\rotn{-90}\def\edgeLen{1*\figScale}
\useasboundingbox ($\figScale*(-0.6,-0.6)$) rectangle ($\figScale*(0.6,0.6)$);
\coordinate (v0) at (0,0);\foreach\a in {0,...,4}{\coordinate (a\a) at ($(v0)+(\a*\rotn-90-\rotn/2:\edgeLen/2)$);};%
\draw[int](a1)--(a2)--(a3)--(a0);\draw[intH](a0)--(a1);\singleLeg{(a0)}{-45};\generalCornerMassive{(a2)}{135};\node[label={[label distance=2*\labelDist]45:{$a$}}] at (a3) {};\node[yshift=-7,xshift=-2,label={[label distance=2*\labelDist]135:{$b$}}] at (a2) {};
\generalCornerMassive{(a1)}{-135};\singleLeg{(a3)}{45};
\node[genAmp] at (a2) {};\node[mhvBarAmp] at (a3) {};\node[genAmp] at (a1) {};\node[mhvAmp] at (a0) {};
\end{tikzpicture}
\hspace{6pt}\right\}\hspace{-50pt}\vspace{6pt}}
which satisfies a residue theorem with precisely the relative signs obtained by evaluating the merger integrands on the soft-collinear residue of (\ref{single_soft_cancel_type2}). The additional contributing terms on this residue, all of which involve $\text{N}^k$MHV boxes with an MHV massless vertex involving leg $a$, cancel as a consequence of the residue theorem involving the two-mass triangle of the opposite chirality, namely,
\vspace{6pt}\eq{\label{grt2}\hspace{-75pt}\bigger{\partial}\!\!\!\left[\rule{0pt}{24pt}\right.\hspace{-8pt}\begin{tikzpicture}[scale=\figScale,baseline=-6.5]\def\rotn{-120}
\coordinate (v0) at (0,0);
\foreach\a in {0,...,3}{\coordinate (a\a) at ($(v0)+(\a*\rotn-30-\rotn/2:\edgeLen/2)$);};
\foreach\a[remember=\a as \la] in {0,...,3}{\ifthenelse{\a=0}{}{\draw[int](a\la)--(a\a)coordinate (e\a) at ($(a\la)!0.5!(a\a)$);}};
\singleLeg{(a0)}{30};
\node[mhvAmp] at (a0) {};
\node[label={[label distance=4*\labelDist]+20:{$a$}}] at (a0) {};
\generalCornerMassive{(a1)}{270};
\node[genAmp] at (a1) {};
\generalCornerMassive{(a2)}{150};
\node[genAmp] at (a2) {};
\node[xshift=-3.0*\figScale,yshift=3.0*\figScale,label={[label distance=4*\labelDist]178:{$b$}}] at (a2) {};
\end{tikzpicture}\left.\hspace{-6pt}\rule{0pt}{28pt}\right]\!\!\!=\!\!\left\{\rule{0pt}{28pt}\hspace{4pt}
\begin{tikzpicture}[scale=\figScale,baseline=-1.5]\def\rotn{-90}\def\edgeLen{1*\figScale}
\useasboundingbox ($\figScale*(-0.6,-0.6)$) rectangle ($\figScale*(0.6,0.6)$);
\coordinate (v0) at (0,0);\foreach\a in {0,...,4}{\coordinate (a\a) at ($(v0)+(\a*\rotn-90-\rotn/2:\edgeLen/2)$);};%
\draw[int](a2)--(a3)--(a0)--(a1);\draw[intH](a1)--(a2);\generalCornerMassive{(a0)}{-45};\generalCornerMassive{(a2)}{135};\node[label={[label distance=2*\labelDist]45:{$a$}}] at (a3) {};\node[xshift=3,label={[label distance=2*\labelDist]-135:{$b$}}] at (a1) {};
\generalCornerMassive{(a1)}{-135};\singleLeg{(a3)}{45};
\node[genAmp] at (a2) {};\node[mhvAmp] at (a3) {};\node[genAmp] at (a1) {};\node[genAmp] at (a0) {};
\end{tikzpicture}\hspace{7pt},\hspace{6pt}
\begin{tikzpicture}[scale=\figScale,baseline=-1.5]\def\rotn{-90}\def\edgeLen{1*\figScale}
\useasboundingbox ($\figScale*(-0.6,-0.6)$) rectangle ($\figScale*(0.6,0.6)$);
\coordinate (v0) at (0,0);\foreach\a in {0,...,4}{\coordinate (a\a) at ($(v0)+(\a*\rotn-90-\rotn/2:\edgeLen/2)$);};%
\draw[int](a2)--(a3)--(a0)--(a1);\draw[intH](a1)--(a2);\generalCornerMassive{(a0)}{-45};\generalCornerMassive{(a2)}{135};\node[label={[label distance=2*\labelDist]45:{$a$}}] at (a3) {};\node[label={[label distance=2*\labelDist]-135:{$b$}}] at (a1) {};
\singleLeg{(a1)}{-135};\singleLeg{(a3)}{45};
\node[genAmp] at (a2) {};\node[mhvAmp] at (a3) {};\node[mhvAmp] at (a1) {};\node[genAmp] at (a0) {};
\end{tikzpicture}\hspace{7pt},\hspace{6pt}
\begin{tikzpicture}[scale=\figScale,baseline=-1.5]\def\rotn{-90}\def\edgeLen{1*\figScale}
\useasboundingbox ($\figScale*(-0.6,-0.6)$) rectangle ($\figScale*(0.6,0.6)$);
\coordinate (v0) at (0,0);\foreach\a in {0,...,4}{\coordinate (a\a) at ($(v0)+(\a*\rotn-90-\rotn/2:\edgeLen/2)$);};%
\draw[int](a2)--(a3)--(a0)--(a1);\draw[intH](a1)--(a2);\generalCornerMassive{(a0)}{-45};\generalCornerMassive{(a1)}{-135};\node[label={[label distance=2*\labelDist]45:{$a$}}] at (a3) {};\singleLeg{(a2)}{135};\singleLeg{(a3)}{45};\node[xshift=3,label={[label distance=2*\labelDist]-135:{$b$}}] at (a1) {};
\node[genAmp] at (a1) {};\node[mhvAmp] at (a3) {};\node[mhvBarAmp] at (a2) {};\node[genAmp] at (a0) {};
\end{tikzpicture}\hspace{7pt},\hspace{6pt}
\begin{tikzpicture}[scale=\figScale,baseline=-1.5]\def\rotn{-90}\def\edgeLen{1*\figScale}
\useasboundingbox ($\figScale*(-0.6,-0.6)$) rectangle ($\figScale*(0.6,0.6)$);
\coordinate (v0) at (0,0);\foreach\a in {0,...,4}{\coordinate (a\a) at ($(v0)+(\a*\rotn-90-\rotn/2:\edgeLen/2)$);};%
\draw[int](a1)--(a2)--(a3)--(a0);\draw[intH](a0)--(a1);\generalCornerMassive{(a0)}{-45};\generalCornerMassive{(a2)}{135};\node[label={[label distance=2*\labelDist]45:{$a$}}] at (a3) {};\node[yshift=-7,xshift=-2,label={[label distance=2*\labelDist]135:{$b$}}] at (a2) {};
\generalCornerMassive{(a1)}{-135};\singleLeg{(a3)}{45};
\node[genAmp] at (a2) {};\node[mhvAmp] at (a3) {};\node[genAmp] at (a1) {};\node[genAmp] at (a0) {};
\end{tikzpicture}\hspace{7pt},\hspace{6pt}
\begin{tikzpicture}[scale=\figScale,baseline=-1.5]\def\rotn{-90}\def\edgeLen{1*\figScale}
\useasboundingbox ($\figScale*(-0.6,-0.6)$) rectangle ($\figScale*(0.6,0.6)$);
\coordinate (v0) at (0,0);\foreach\a in {0,...,4}{\coordinate (a\a) at ($(v0)+(\a*\rotn-90-\rotn/2:\edgeLen/2)$);};%
\draw[int](a1)--(a2)--(a3)--(a0);\draw[intH](a0)--(a1);\generalCornerMassive{(a0)}{-45};\generalCornerMassive{(a2)}{135};\node[label={[label distance=2*\labelDist]45:{$a$}}] at (a3) {};\node[yshift=-7,xshift=-2,label={[label distance=2*\labelDist]135:{$b$}}] at (a2) {};
\singleLeg{(a1)}{-135};\singleLeg{(a3)}{45};
\node[genAmp] at (a2) {};\node[mhvAmp] at (a3) {};\node[mhvAmp] at (a1) {};\node[genAmp] at (a0) {};
\end{tikzpicture}\hspace{7pt},\hspace{6pt}
\begin{tikzpicture}[scale=\figScale,baseline=-1.5]\def\rotn{-90}\def\edgeLen{1*\figScale}
\useasboundingbox ($\figScale*(-0.6,-0.6)$) rectangle ($\figScale*(0.6,0.6)$);
\coordinate (v0) at (0,0);\foreach\a in {0,...,4}{\coordinate (a\a) at ($(v0)+(\a*\rotn-90-\rotn/2:\edgeLen/2)$);};%
\draw[int](a1)--(a2)--(a3)--(a0);\draw[intH](a0)--(a1);\singleLeg{(a0)}{-45};\generalCornerMassive{(a2)}{135};\node[label={[label distance=2*\labelDist]45:{$a$}}] at (a3) {};\node[yshift=-7,xshift=-2,label={[label distance=2*\labelDist]135:{$b$}}] at (a2) {};
\generalCornerMassive{(a1)}{-135};\singleLeg{(a3)}{45};
\node[genAmp] at (a2) {};\node[mhvAmp] at (a3) {};\node[genAmp] at (a1) {};\node[mhvBarAmp] at (a0) {};
\end{tikzpicture}
\hspace{6pt}\right\}\hspace{-50pt}\vspace{6pt}}
Of course, for a given choice of $k$, not all of the on-shell functions appearing in the general residue theorems (\ref{grt1}) and (\ref{grt2}) will be non-zero. For example, for N${}^{(k=1)}$MHV amplitudes the latter identity (\ref{grt2}) will involve only two terms. In fact, even the cancellation in (\ref{single_soft_cancel_type1}) can be regarded as another instance of a two-term residue theorem for MHV amplitude integrands. The remaining non-soft-but-collinear divergent regions can also be dealt with using almost the exact same argument, as we shall now demonstrate.

\vspace{-0pt}\subsubsection{non-Soft-but-Collinear $\logdiv$-Divergent Regions}\label{subsubsec:other_log1_div_regions}\vspace{-0pt}
%
The final class of divergences which must cancel to prove the local finiteness of the ratio function are those associated to collinear-but-not-soft singularities in \emph{both} loops. We must demonstrate the cancellation of all locally divergent regions in loop momentum space parametrized as 
\eq{\Big\{\ell_1=\alpha^\star\,x_a+(1-\alpha^\star)\,x_{a+1},\,\, \ell_2=\beta^\star\, x_a+(1-\beta^\star)\,x_{a+1}\,,\,\,\x{\ell_1(\alpha^\star)}{b}=\x{\ell_2(\beta^\star)}{c}=0\Big\}\,.}
Here, $\alpha^\star,\beta^\star$ are the solutions to cutting two additional (non-adjacent) propagators $\x{\ell_1}{b}$ and $\x{\ell_2}{c}$ in the two loops, respectively. In this case, neither loop can access the soft-collinear region, and the relevant mergers were given in (\ref{coll_but_not_soft_ints}). It is once again useful to organize the cancellations according to whether we `collect' on the MHV loop or the $\text{N}^k$MHV loop. 

First, consider the mergers which contribute in the divergent region where both loops are collinear to leg $a$, and the MHV loop is fully localized by setting $\x{\ell_2}{c}=0$. For each $\text{N}^k$MHV box there are two MHV boxes with support on this residue, and the cancellation is pairwise,
\eq{\underset{\substack{{\ell_1,\ell_2\in\text{coll}(a)} \\ {\x{\ell_2}{c}=0}}}{\Res}\mathcal{A}_{n,\mathrm{fin}}^{(k),1}\merge\mathcal{A}_{n,\mathrm{fin}}^{(0),1}\bigger{\supset}\fig{-22.5pt}{1}{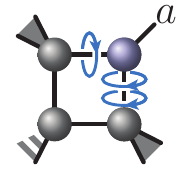}\merge\left(\fig{-22.5pt}{1}{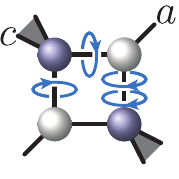}+\fig{-22.5pt}{1}{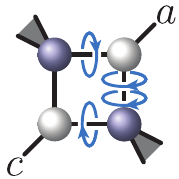}\right)=0.}
Here, we can see these two residues cancel as functions of the remaining loop degree of freedom, even prior to localizing e.g. $\x{\ell_1}{b}=0$. If, on the other hand, we collect on (a collinear-but-not-soft pole of) the $\text{N}^k$MHV loop, the contributing on-shell functions are:\\[-12pt] 
\begin{align}\hspace{-40pt}\underset{\substack{{\ell_1,\ell_2\in\text{coll}(a)} \\ {\x{\ell_1}{b}=\x{\ell_2}{c}=0}}}{\Res}\mathcal{A}_{n,\mathrm{fin}}^{(k),1}\merge\mathcal{A}_{n,\mathrm{fin}}^{(0),1}\,\bigger{\supset}&\hspace{-8.5pt}\phantom{+}\Bigg(\fig{-22.5pt}{1}{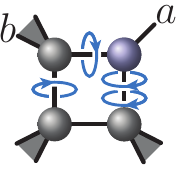}+\fig{-22.5pt}{1}{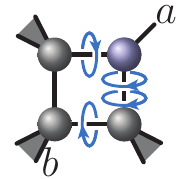}+\fig{-22.5pt}{1}{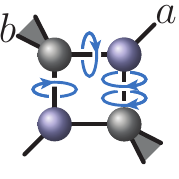} \\ &\!\!+\fig{-22.5pt}{1}{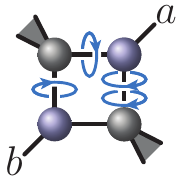}+\fig{-22.5pt}{1}{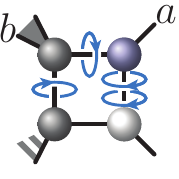}+\fig{-22.5pt}{1}{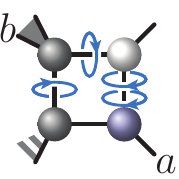}\Bigg)\merge\fig{-22.5pt}{1}{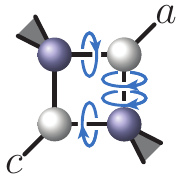}=0\hspace{-24pt}\nonumber\end{align}
(where we are implicitly summing over \emph{all} possible factorizations with the requisite pole structure). Once again, these combine to give a residue theorem generated by (\ref{grt2}), cancelling all of these local divergences.\\

We have therefore exhausted our classification of local divergences in two loop amplitude integrands, showing that each possibly divergent region of loop momentum space is cancelled between terms appearing in the expression for the ratio function integrand given in (\ref{ratio_function_explosion}). Thus, we have proven---for all multiplicity and arbitrary N$^{k}$MHV degree, that two-loop ratio functions in planar sYM are \emph{locally finite}.

%
\vspace{-0pt}\section{Conclusions and Future Directions}\label{sec:conclusions}\vspace{-0pt}

In this paper, we have investigated the infrared behavior of the two-loop ratio function in planar, maximally supersymmetric ($\mathcal{N}=4$) Yang-Mills theory (sYM). By introducing integrand-level tests to locally diagnose two-loop infrared-finiteness, we have identified the correspondence between particular soft and collinear residues at the integrand-level, and the maximal degree of divergence which arise upon integration in some regularization scheme. We have shown that the particular combination of infrared-divergent amplitudes in the two-loop ratio function is locally finite by explicitly demonstrating the cancellation of all divergences present in the local integrand representation of \cite{Bourjaily:2015jna}. 

Having demonstrated the finiteness of the ratio function, there is an immediate follow-up to this work: namely, finding a \emph{new} representation of ratio functions which makes their finiteness \emph{manifest} at the integrand-level. With such, the ratio function would be possible to compute without regularization---perhaps making the preservation of the simplicities of the integrand (such as dual conformal invariance) more manifest after integration.

Because local finiteness is a homogeneous constraint on any loop integrand, it is relatively straightforward to \emph{construct} the space of locally-finite box-power-counting two loop integrands for any multiplicity (as a subspace of all integrands with box power-counting). However, it is by no means guaranteed that a basis for this space of integrands exists which is compatible with dual-conformality, purity, ease of integration, or even stability in form with increasing multiplicity. We suspect that a basis of pure, dual-conformal, locally finite integrands exist for all multiplicity at two loops analogous to the one described at one-loop in \cite{ArkaniHamed:2010gh}; however, we must leave this search to future work. 

Although our proof of local finiteness of the ratio function at two loops relied heavily on the representation (\ref{ratio_function_explosion}) derived in \cite{Bourjaily:2019gqu}, and involved checking a number of specific case, there is reason to suspect that a more powerful, all-loop argument may follow from a different line of reasoning. Indeed, in \cite{Drummond:2007au} it was shown that the divergences (at the integrand-level) of the \emph{logarithm} of the amplitude in the Wilson-loop picture were entirely captured by the contributions from cusps; such an understanding of the local (in loop-momentum-space) structure responsible for infrared divergences is something that would be fruitful to explore for a broader class of infrared-(or ultraviolet-)finite observables. 

\vspace{6pt}
\section*{Acknowledgements}%
\vspace{-4pt}
\noindent The authors are grateful for fruitful conversations with Enrico Herrmann, Gregory Korchemsky, Radu Roiban, and Jaroslav Trnka. This project has been supported by an ERC Starting Grant \mbox{(No.\ 757978)} and a grant from the Villum Fonden \mbox{(No.\ 15369)}.

%

\newpage
\providecommand{\href}[2]{#2}\begingroup\raggedright\endgroup

\end{document}